\documentclass[12pt]{iopart}

\usepackage{graphicx}
\usepackage{bm}
\usepackage{color}
\usepackage{multirow}
\usepackage{iopams}

\newcommand{\bra}[1]{\left\langle #1 \right|}
\newcommand{\ket}[1]{\left| #1 \right\rangle}

\begin{document}

\title{Coexistence of nuclear shapes: self-consistent mean-field and beyond}
\author{Z. P. Li$^{1}$, T. Nik\v si\' c$^{2}$, D. Vretenar$^{2}$}
\address{$^{1}$School of Physical Science and Technology, Southwest University, 400715 Chongqing, China}
\address{$^{2}$Physics Department, Faculty of Science, University of Zagreb,
10000 Zagreb, Croatia}

\begin{abstract}
A quantitative analysis of the evolution of nuclear shapes and shape phase transitions, including regions of 
short-lived nuclei that are becoming accessible in experiments at radioactive-beam facilities, necessitate 
accurate modeling of the underlying nucleonic dynamics. Important theoretical advances have 
recently been made in studies of complex shapes and the corresponding excitation spectra and electromagnetic 
decay patterns, especially in the ``beyond mean-field" framework based on nuclear density functionals. Interesting 
applications include studies of shape evolution and coexistence in $N=28$ isotones, the structure of
lowest $0^+$ excitations in deformed $N \approx 90$ rare-earth nuclei, and quadrupole and octupole shape transitions 
in thorium isotopes. 
\end{abstract}

\maketitle

%============================================================
%  Section 1
\section{\label{secI}Introduction}
%============================================================
One of the most studied phenomenon in low-energy nuclear physics, both experimentally and 
theoretically, is the way nucleonic matter organises itself to support a variety of 
shapes observed in finite nuclei. The occurrence of different shapes, shape coexistence, 
and shape transitions have their origin in the evolution of single-nucleon shell structure 
with nuclear deformation, angular momentum, temperature and number of valence nucleons. 
The manifestation of shells is a generic property of finite fermion systems and shell closures, 
in particular, are characteristic of the confining single-particle potential. When nucleons 
completely fill a major shell (single or doubly closed-shell nuclei), the relatively large energy gap 
to the next shell stabilises a spherical shape, whereas long-range correlations between 
valence nucleons in open-shell nuclei drive the nucleus toward deformed (quadrupole, octupole) 
equilibrium shapes. 

Coexistence of different shapes in a single nucleus, and shape (phase) transitions as a function of 
nucleon number, present universal phenomena that occur in light, medium-heavy, heavy and 
superheavy nuclei, and reflect the organisation of nucleons in finite nuclei \cite{HW.11,BN.96,SP.08,CJC.10}. 
A unified description of shape evolution and shape coexistence over the entire chart of nuclides 
necessitates a universal theory framework that can be applied to different mass regions. 
Nuclear energy density functionals (EDF) provide an economic, global and accurate 
microscopic approach to nuclear structure that can be extended  
from relatively light systems to superheavy nuclei, and from the
valley of $\beta$-stability to the particle drip-lines \cite{BHR.03,VALR.05,LNP.641,SR.07}. 
This is particularly important for extrapolations to regions far from stability where not enough 
data are available to determine the parameters of a more local approach such as, for 
instance, the interacting shell model. 

The basic implementation of the EDF framework is in terms of
self-consistent mean-field (SCMF) models, in which an EDF is constructed
as a functional of one-body nucleon density matrices that correspond to
a single product state. Nuclear SCMF models
effectively map the many-body problem onto a one-body
problem, and the exact EDF is approximated by simple   
functionals of powers and gradients of ground-state nucleon densities and currents. 
Some of the advantages of the EDF approach are apparent already at the 
SCMF level: an intuitive interpretation of mean-field results in terms of intrinsic 
shapes and single-nucleon states, and the possibility to formulate structure models 
in the full model space of occupied states, with no distinction between core and 
valence nucleons. 

Quantitative studies of low-energy structure phenomena related to shell 
evolution and coexistence usually start from a constrained Hartree-Fock
plus BCS (HFBCS), or Hartree-Fock-Bogoliubov (HFB) calculation of the
deformation energy surfaces with mass multipole moments as
constrained quantities. When based on microscopic EDFs or effective
interactions, such calculations comprise many-body correlations 
related to the short-range repulsive inter-nucleon interaction and 
long-range correlations mediated by nuclear resonance modes.
The result are static symmetry-breaking product many-body states. 
The basic idea of a deformation energy surface is that, even though the quantum  
many-body system is determined by a very large number of microscopic states, 
these can be organised in a collection of basins that are robust to small external 
perturbations \cite{Laughlin.00}. The basins can be structurally distinct and 
distant from each other, but they can occur at comparable energies. 

The constrained SCMF method, however, produces semi-classical 
deformation energy surfaces. The static nuclear mean-field is characterised by the 
breaking of symmetries of the underlying Hamiltonian -- translational, rotational, particle 
number and, therefore, includes important static correlations, e.g. deformations and 
pairing. To calculate excitation spectra and electromagnetic transition rates
it is necessary to extend the SCMF scheme to include collective correlations that
arise from symmetry restoration and fluctuations around the mean-field
minima. Collective correlations are sensitive to shell effects, display
pronounced variations with particle number and, therefore, cannot be
incorporated in a universal EDF but rather require an explicit 
treatment \cite{BHR.03,NVR.11,LNP.879}. 

On the second level of implementation of nuclear EDFs that takes into
account collective correlations through the restoration of broken symmetries
and configuration mixing of symmetry-breaking product states, the
many-body energy takes the form of a functional of all transition
density matrices that can be constructed from the chosen set of
product states. This set is chosen to restore symmetries or/and to
perform a mixing of configurations that correspond to specific collective
modes using, for instance, the (quasiparticle) random-phase approximation
(QRPA) or the Generator Coordinate Method (GCM) \cite{RS.80}.
The latter includes correlations related to finite-size fluctuations in a collective
degree of freedom and presents the most effective approach for
configuration mixing calculations, with multipole moments used as coordinates 
that generate the intrinsic wave functions.

Many interesting phenomena related to shell evolution have been investigated 
over the last decade by employing GCM configuration mixing of angular-momentum 
and particle-number projected states based on energy density functionals or effective interactions,
but the extension of this method to non-axial shapes and/or heavy nuclei still presents 
formidable conceptual and computational challenges \cite{LNP.879,BH.08,RE.10,Yao.10,BABH.14}. 
In an alternative approach to nuclear collective dynamics that restores rotational symmetry and 
allows for fluctuations around mean-field minima, a collective Hamiltonian
can be formulated, with deformation-dependent parameters determined by
self-consistent mean-field calculations \cite{RG.87,PR.09}. 
The dynamics of the collective Hamiltonian is governed by the
vibrational inertial functions and the moments of inertia \cite{GR.80}, 
and these functions are
determined by the microscopic nuclear energy
density functional and the effective interaction in the pairing channel. 
Five-dimensional collective Hamiltonian models for quadrupole vibrational and rotational
degrees of freedom, with parameters determined by constrained triaxial SCMF calculations based 
on the Gogny effective interaction \cite{Delaroche10}, the Skyrme density functional \cite{PR.09}, and 
relativistic density functionals \cite{NVR.11}, have been developed over the last decade and 
applied in a number of studies of structure phenomena related to shape coexistence and 
shape transitions. 

The present study is based on the framework of relativistic energy density functionals, extended 
to include the treatment of collective correlations using the collective 
Hamiltonian model \cite{NVR.11}. To emphasise the universality of the EDF approach, all 
illustrative calculations performed in this study, from relatively light 
systems to very heavy nuclei, have been carried out using a single energy density functional --
DD-PC1 \cite{Nik.08}. Starting from microscopic nucleon self-energies in nuclear matter, 
and empirical global properties of the nuclear matter equation of state, the coupling parameters 
of DD-PC1 were fine-tuned to the experimental masses of a set of 64 deformed nuclei in the 
mass regions $A \approx 150 -180$ and $A \approx 230 - 250$. The functional has been further 
tested in a number of mean-field and beyond-mean-field calculations in different mass 
regions. For the examples considered here, pairing correlations have been taken into account 
by employing an interaction that is separable in momentum space, and is completely determined 
by two parameters adjusted to reproduce the empirical bell-shaped pairing gap in symmetric 
nuclear matter \cite{Tian_PLB.09}. For the details of the particular implementation of the 
EDF-based collective Hamiltonian used in the present study, 
we refer the reader to Ref.~\cite{NVR.11}. 
%============================================================
%  Section 2
\section{\label{secII}Beyond the relativistic mean-field approximation: collective correlations}
%============================================================

For a self-consistent description of collective excitation spectra 
and electromagnetic transition rates, the framework of (relativistic) 
energy density functionals has to be extended to take into account 
collective correlations in relation to  
restoration of broken symmetries and fluctuations in collective coordinates. 
Both types of correlations can be included simultaneously by mixing symmetry-projected states 
corresponding to different values of chosen collective coordinates. 
The most effective approach for configuration mixing calculations is the generator 
coordinate method (GCM), with multipole moments used as coordinates that generate the 
intrinsic wave functions.
The GCM is based on the assumption
that, starting from a set of mean-field states $\ket{\Phi (q)}$ that 
depend on a collective coordinate $q$, one can build
approximate eigenstates of the nuclear Hamiltonian:
\begin{equation}
\ket{\Psi_\alpha} =  \int d q {f_\alpha(q)\ket{\Phi (q)}}\;.
\label{GCM-state}
\end{equation}
Here the basis states $\ket{\Phi (q)}$ are Slater determinants of single-nucleon states 
generated by constrained SCMF calculations. Several advanced implementations of the GCM have been 
developed recently, fully based on the microscopic EDF framework. For the   
relativistic SCMF approach, in particular, the most advanced model performs 
configuration mixing of angular-momentum and particle-number
projected wave functions generated by constraints on quadrupole deformations \cite{Yao.10,Yao.14}
\begin{equation}
|JNZ;\alpha \rangle = \int{dq\sum_{K}{f_{\alpha}^{JK} \hat{P}^J_{MK} \hat{P}^N\hat{P}^Z |q\rangle } },
\end{equation}

\noindent
where $\alpha = 1,2,\dots$ denotes different collective states for a given angular momentum $J$,
and $|q\rangle \equiv |\beta, \gamma\rangle$ denotes a set of intrinsic SCMF states with deformation parameters
$(\beta,\gamma)$. 
$\hat{P}_{MK}^J$ is the angular momentum projection operator, and 
the operators $\hat{P}^N$ and $\hat{P}^Z$ project onto states with 
good neutron and proton number, respectively. 
This implementation is equivalent to a seven-dimensional GCM calculation, mixing all five degrees of freedom of the 
quadrupole operator and the gauge angles for protons and neutrons. 
The weight functions $f^{JK}_{\alpha}(q)$ in the collective wave function 
are determined from the variational equation:
\begin{equation}
 \delta E^{J} =
 \delta \frac{\bra{\Psi_\alpha^{JM}} \hat{H} \ket{\Psi_\alpha^{JM}}}
            {\bra{\Psi_\alpha^{JM}}\Psi_\alpha^{JM}\rangle} = 0 \; ,
\label{variational}
\end{equation}
that is, by requiring that the expectation value of the Hamiltonian is
stationary with respect to an arbitrary variation $\delta
f_{\alpha}^{JK}$. This leads to the  Hill-Wheeler-Griffin (HWG)
integral equation:
\begin{equation}
 \label{HWEq}
 \int dq^\prime\sum_{K^\prime\geq 0}
 \left[\mathcal{H}^J_{KK^\prime}(q,q^\prime)
 - E^J_\alpha\mathcal{N}^J_{KK^\prime}(q,q^\prime)\right]
  f^{JK^\prime}_\alpha(q^\prime)=0,
\end{equation}
 where $\mathcal{H}$ and $\mathcal{N}$ are the projected GCM
 kernel matrices of the Hamiltonian and the norm, respectively \cite{BH.08,RE.10,Yao.10,Yao.14}.

Multidimensional GCM calculations involve a number of technical and computational issues \cite{LNP.879,Yao.14,BABH.14}, 
that have so far impeded systematic applications to medium-heavy and heavy nuclei. 
Collective dynamics can also be described using an alternative method in which 
a collective Hamiltonian is constructed, with deformation-dependent parameters determined from
microscopic SCMF calculations \cite{RG.87,PR.09}. 
The collective Hamiltonian can be derived in
the Gaussian overlap approximation (GOA)~\cite{RS.80} to the full
multi-dimensional GCM. With the assumption that the GCM overlap kernels can be
approximated by Gaussian functions, the local expansion of the
kernels up to second order in the non-locality transforms the
GCM Hill-Wheeler equation into a second-order differential equation -
the Schr\"odinger equation for the collective Hamiltonian. 
For instance, in the case of quadrupole degrees of freedom:
\begin{equation}
\label{hamiltonian-quad}
\hat{H}_{\rm coll} = \hat{T}_{\rm{vib}}+\hat{T}_{\rm{rot}}
              +V_{\rm{coll}} \; ,
\end{equation}
where the vibrational kinetic energy is parameterized by the the mass parameters
$B_{\beta\beta}$, $B_{\beta\gamma}$, $B_{\gamma\gamma}$
%------------------------------------------------------------------------------------
\begin{eqnarray}
\hat{T}_{\textnormal{vib}} =-\frac{\hbar^2}{2\sqrt{wr}}
   \left\{\frac{1}{\beta^4}
   \left[\frac{\partial}{\partial\beta}\sqrt{\frac{r}{w}}\beta^4
   B_{\gamma\gamma} \frac{\partial}{\partial\beta}
   - \frac{\partial}{\partial\beta}\sqrt{\frac{r}{w}}\beta^3
   B_{\beta\gamma}\frac{\partial}{\partial\gamma}
   \right]\right.
   \nonumber \\
   +\frac{1}{\beta\sin{3\gamma}}\left.\left[
   -\frac{\partial}{\partial\gamma} \sqrt{\frac{r}{w}}\sin{3\gamma}
      B_{\beta \gamma}\frac{\partial}{\partial\beta}
    +\frac{1}{\beta}\frac{\partial}{\partial\gamma} \sqrt{\frac{r}{w}}\sin{3\gamma}
      B_{\beta \beta}\frac{\partial}{\partial\gamma}
   \right]\right\} \; ,
 \end{eqnarray}
 %------------------------------------------------------------------------------------
the three moments of inertia $\mathcal{I}_k$ determine the 
rotational kinetic energy 
%------------------------------------------------------------------------------------
\begin{equation}
\hat{T}_{\textnormal{\textnormal{\textnormal{rot}}}} = 
\frac{1}{2}\sum_{k=1}^3{\frac{\hat{J}^2_k}{\mathcal{I}_k}} \; ,
\end{equation}
%------------------------------------------------------------------------------------
and $V_{\rm{coll}}$ is the collective potential that includes zero-point
energy (ZPE) corrections. $w$ and $r$ are products of mass parameters 
and moments of inertia, respectively, that specify the volume element in  
collective space \cite{Nik.09}. 
The self-consistent mean-field solution for the 
single-quasiparticle energies and wave functions for
the entire energy surface, as functions of the quadrupole
deformations $\beta$ and $\gamma$, provide the microscopic input for 
calculation of the mass parameters, moments of inertia and the collective 
potential. The Hamiltonian describes quadrupole vibrations,
rotations, and the coupling of these collective modes.
 
The dynamics of the collective Bohr Hamiltonian is governed by the
vibrational inertial functions and the moments of inertia \cite{GR.80}.
For these quantities either the GCM-GOA 
or the adiabatic approximation to the time-dependent HFB (ATDHFB) expressions 
(Thouless-Valatin masses) can be used. The Thouless-Valatin masses
have the advantage that they also include the time-odd components of
the mean-field potential and, in this sense, the full dynamics of a
nuclear system. In the GCM approach these components can only be included if,
in addition to the coordinates  $q_i$, the corresponding
canonically conjugate momenta $p_i$ are also taken into account. 
In many applications, including the collective Hamiltonians considered in the present study, a further
simplification is thus introduced in terms of cranking formulas,
i.e. the perturbative limit for the Thouless-Valatin masses, and the
corresponding expressions for ZPE corrections \cite{Nik.09}. In the present implementation of the 
collective Hamiltonian model the moments of inertia and 
mass parameters do not include the contributions of time-odd mean-fields (the so 
called dynamical rearrangement contributions) and, to a certain extent, this breaks 
the self-consistency of the approach \cite{Hin.12}. 

The diagonalization of the collective Hamiltonian gives the 
energy spectrum $E_\alpha^I$ and the corresponding eigenfunctions
\begin{equation}
\label{wave-coll}
\Psi_\alpha^{IM}(\beta,\gamma,\Omega) =
  \sum_{K\in \Delta I}
           {\psi_{\alpha K}^I(\beta,\gamma)\Phi_{MK}^I(\Omega)}\; , 
\end{equation}
that are used to calculate various observables, for instance the E2 reduced
transition probabilities. The shape of a nucleus can be characterized in a qualitative way
by the expectation values of invariants $\beta^2$, $\beta^3\cos{3\gamma}$, as well as their
combinations.

Nuclear excitations characterised by quadrupole and octupole vibrational and rotational degrees of
freedom can be simultaneously described by considering quadrupole and octupole collective coordinates
that specify the surface of a nucleus $R=R_0\left[1+\sum_\mu{\alpha_{2\mu}Y^*_{2\mu}
  + \sum_\mu{\alpha_{3\mu}Y_{3\mu}^*} } \right]$. In addition, when axial symmetry is imposed, 
  the collective coordinates
  can be parameterized in terms of two deformation parameters $\beta_2$ and $\beta_3$, and three Euler
  angles  $ \Omega\equiv(\phi,\theta,\psi)$. After quantization the collective Hamiltonian takes the form
\begin{eqnarray}
{\hat H}_{\rm coll} &= -\frac{\hbar^2}{2\sqrt{w{\cal I}}}
              \left[\frac{\partial}{\partial\beta_2}\sqrt{\frac{{\cal I}}{w}}B_{33}\frac{\partial}{\partial\beta_2}
                      -\frac{\partial}{\partial\beta_2}\sqrt{\frac{{\cal I}}{w}}B_{23}\frac{\partial}{\partial\beta_3} 
            -\frac{\partial}{\partial\beta_3}\sqrt{\frac{{\cal I}}{w}}B_{23}\frac{\partial}{\partial\beta_2}\right.\nonumber\\
&
                       \left. +\frac{\partial}{\partial\beta_3}\sqrt{\frac{{\cal I}}{w}}B_{22}\frac{\partial}{\partial\beta_3}\right.
           +\frac{\hat{J}^2}{2{\cal I}}+{V}(\beta_2, \beta_3),
   \label{eq:CH2}
\end{eqnarray}
where the mass parameters $B_{22}$, $B_{23}$ and $B_{33}$, and the moment of inertia
$\mathcal{I}$, are functions of the quadrupole  $\beta_2$ and octupole $\beta_3$ deformations.
$w=B_{22}B_{33}-B_{23}^2$. 

Just as in the case of the quadrupole five-dimensional collective Hamiltonian, the moments of inertia 
are
calculated from the Inglis-Belyaev formula:
\begin{equation}
\label{Inglis-Belyaev}
\mathcal{I}= \sum_{i,j}{\frac{| \langle ij |\hat{J} | \Phi \rangle |^2}{E_i+E_j}}\;,
\end{equation}
where $\hat{J}$ is the angular momentum along the axis perpendicular to the symmetric axis, 
the summation runs over
proton and neutron quasiparticle states
$|ij\rangle=\beta^\dagger_i\beta^\dagger_j|\Phi\rangle$, and
$|\Phi\rangle$ represents the quasiparticle vacuum. The quasiparticle
energies $E_i$ and wave functions are determined by SCMF calculations of 
deformation energy surfaces with constraints on the quadrupole and octupole 
deformation parameters. 
The mass parameters associated with the collective 
coordinates $q_2=\langle\hat{Q}_{2}\rangle$ and $q_3=\langle\hat{Q}_{3}\rangle$
are calculated in the cranking approximation, as well as the vibrational and rotational   
zero-point energy corrections to the collective energy surface \cite{GG.79}. 
%============================================================
%  Section 3
\section{\label{secIII}Evolution of shapes and coexistence in $\bm{N=28}$ isotones}
%============================================================
Nuclei with closed major proton and/or neutron shells are usually characterised by spherical 
equilibrium shapes. However, this is not necessarily the case for nuclei away from 
the $\beta$-stability line in which energy spacings between single-particle levels 
can undergo considerable changes 
with the number of neutrons and/or protons. This can result in
reduced spherical shell gaps, modifications of shell structure,
and in some cases spherical magic numbers may disappear.
The reduction of a spherical shell closure is associated with the
occurrence of deformed ground states and, in a number of cases,
with the phenomenon of shape coexistence. Because of the low density of 
single-particle states close to the Fermi surface, in relatively light nuclei coexistence 
occurs only when proton and neutron density distributions favour different 
equilibrium deformations, e.g. prolate {\em vs} oblate shapes. 
Here we consider the well known example of neutron-rich $N=28$ isotones, which 
exhibit rapid shape variations and shape coexistence. 

\begin{figure}[]
\begin{center}
\begin{tabular}{cc}
\includegraphics[scale=0.42]{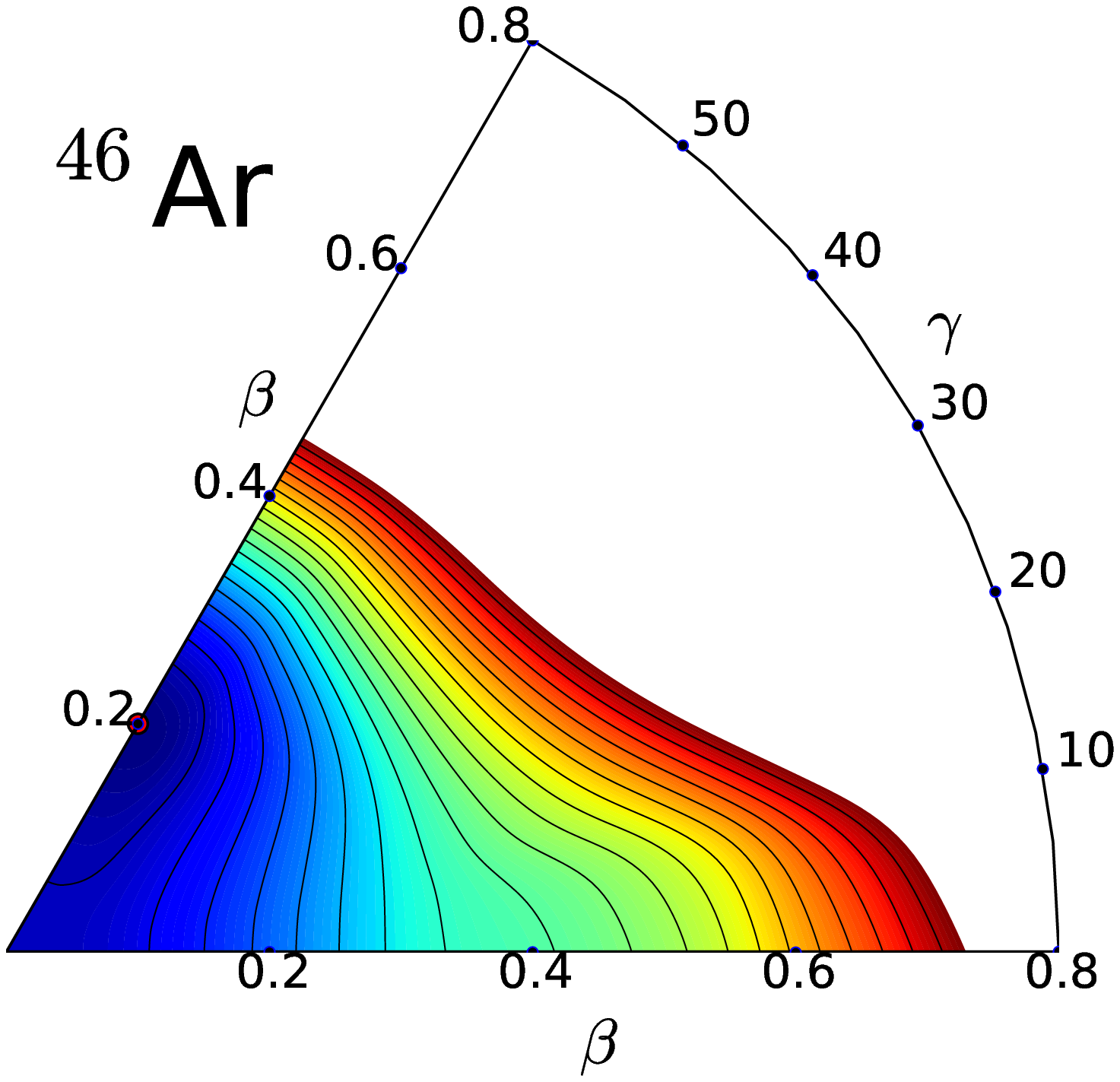}&
\hspace{-1cm}\includegraphics[scale=0.42]{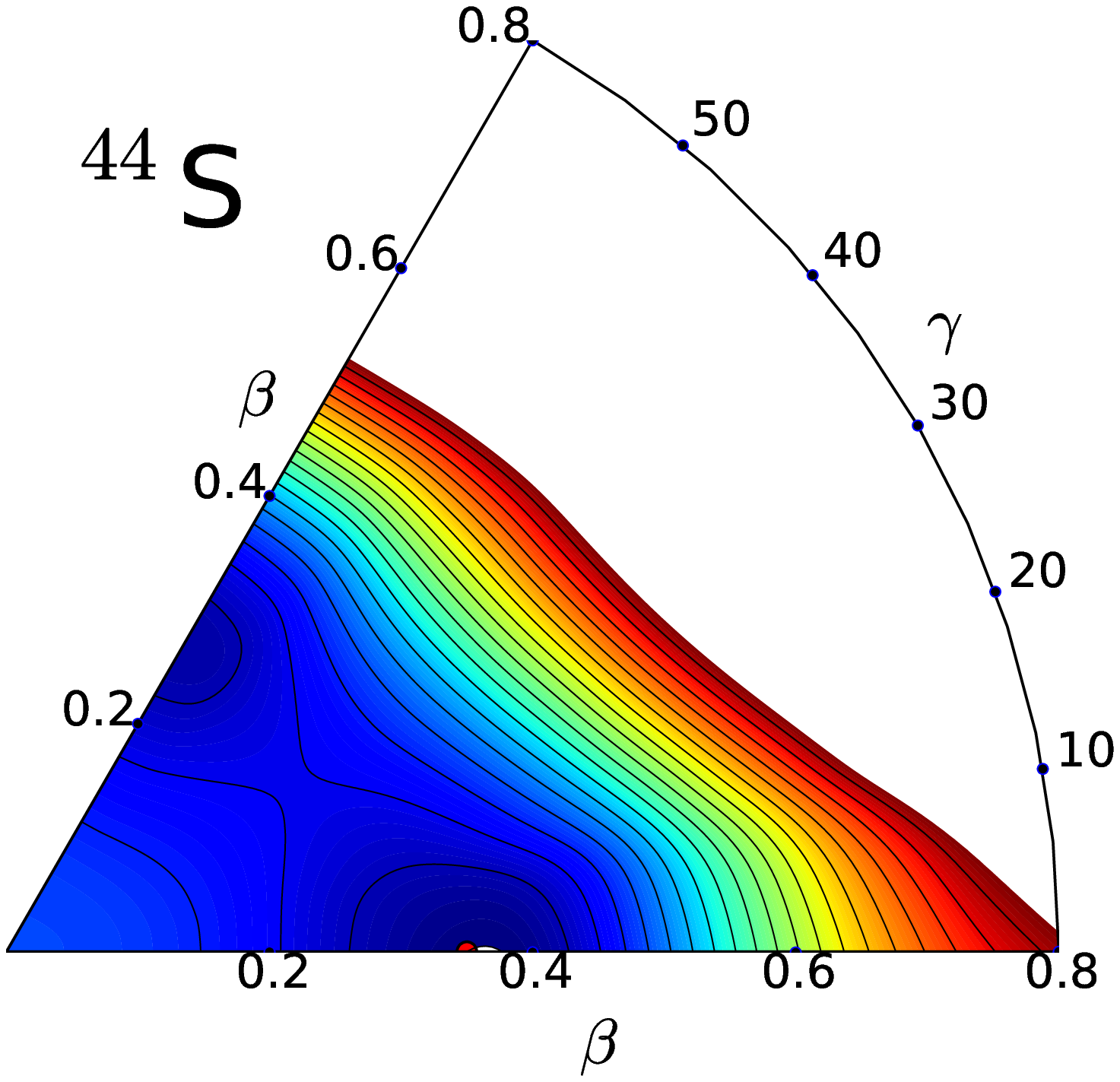} \\
\includegraphics[scale=0.42]{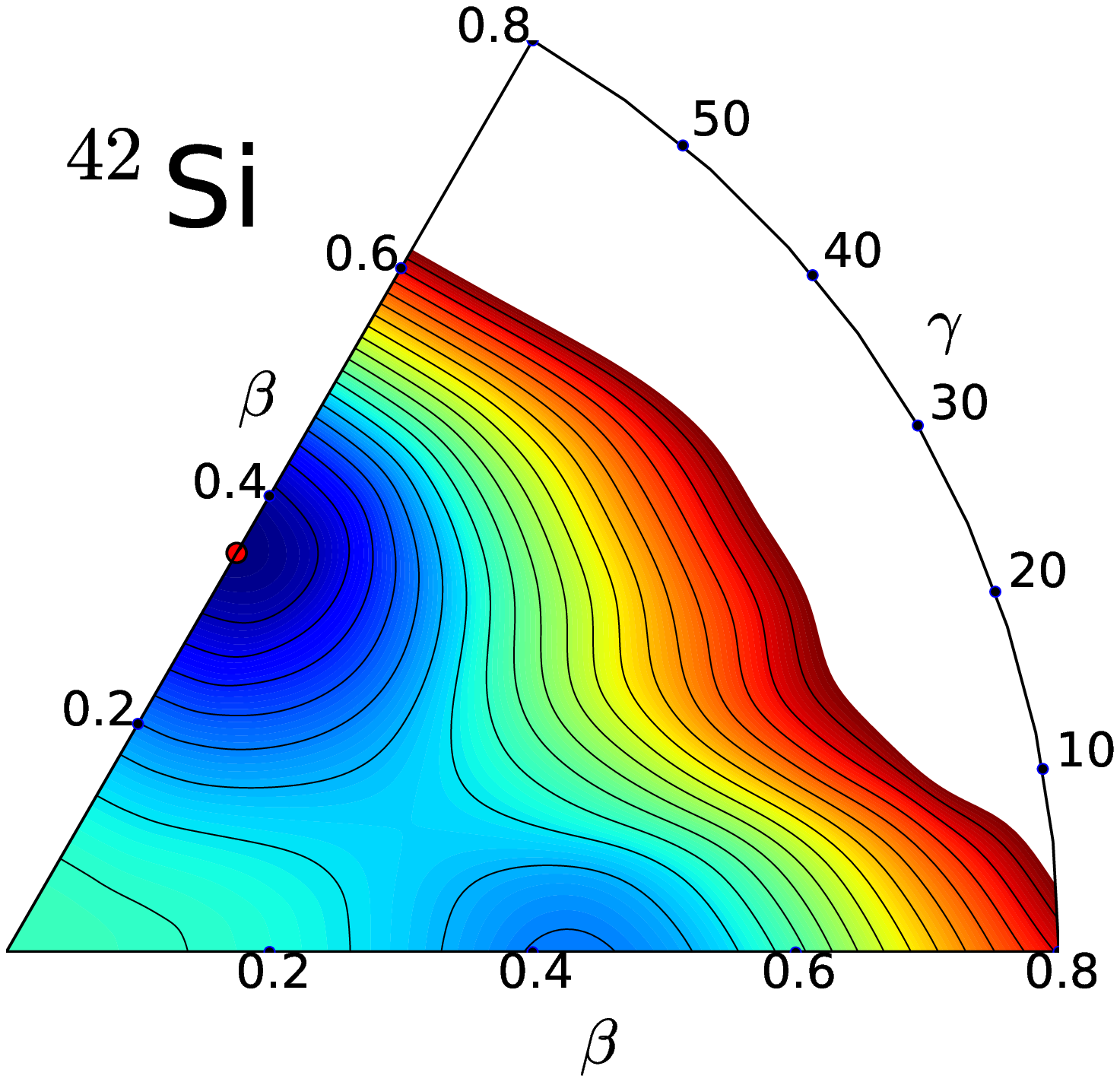}&
\hspace{-1cm}\includegraphics[scale=0.42]{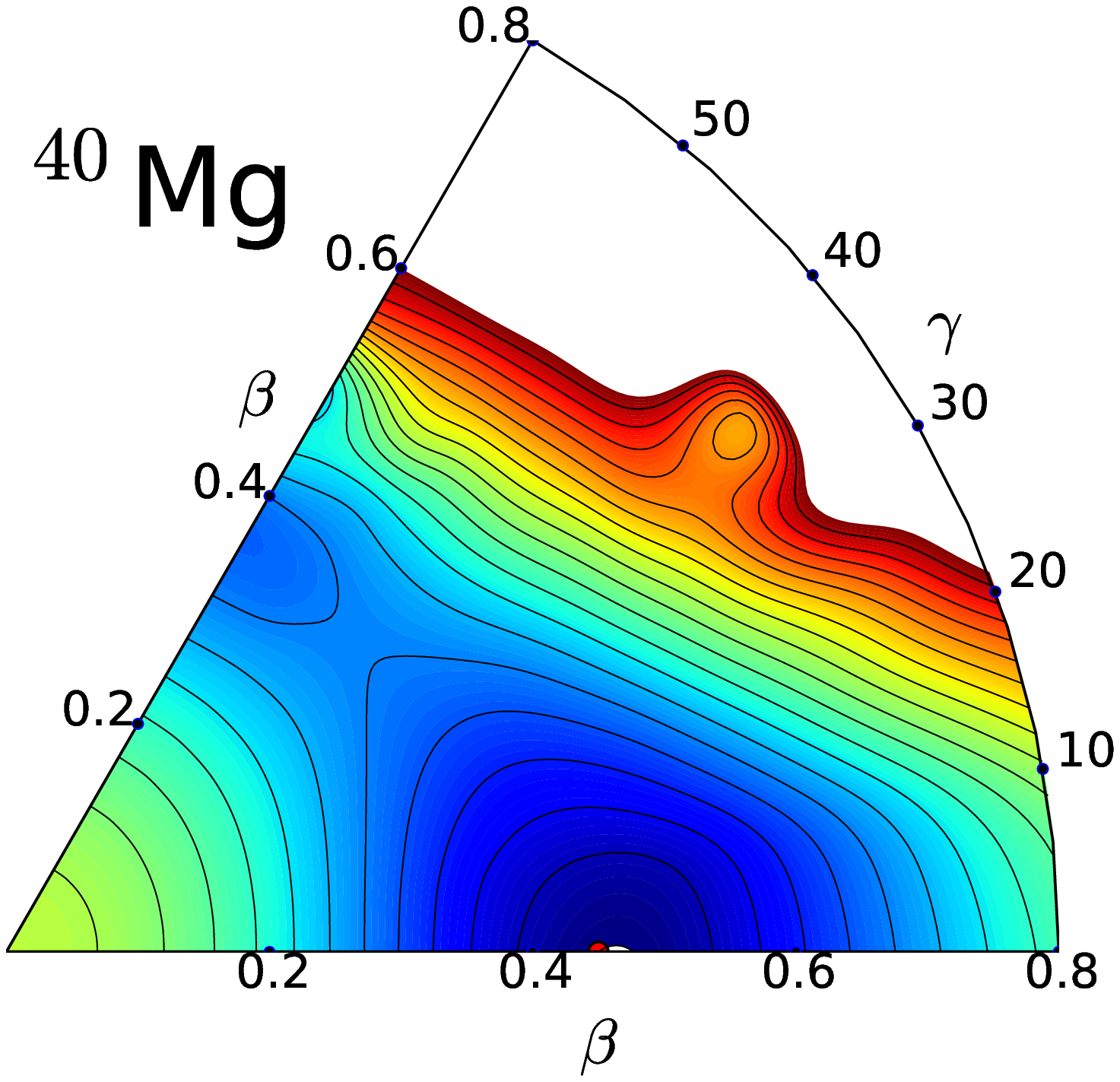} 
\end{tabular}
\caption{\label{fig:n28_pes} Self-consistent triaxial
quadrupole constrained energy surfaces of $N=28$ isotones in the $\beta-\gamma$ plane
($0\le \gamma \le 60^0$). For each nucleus energies are normalized with respect to the
binding energy of the global minimum. The contours join points on
the surface with the same energy, and the spacing between 
neighbouring contours is 0.5 MeV.}
\end{center}
\end{figure}

In Fig.~\ref{fig:n28_pes} we plot the self-consistent triaxial 
quadrupole deformation energy surfaces of $N=28$ isotones \cite{Li.11}. 
The equilibrium shape of the doubly-magic nucleus $^{48}$Ca is, of course, 
spherical but the $N=28$ spherical shell is strongly reduced in the isotones 
with a smaller number of protons. This leads to rapid transitions between deformed equilibrium 
shapes and shape coexistence in $^{44}$S. 
The energy surface of $^{46}$Ar is 
soft both in $\beta$ and $\gamma$ directions, with a shallow extended minimum
along the oblate axis. Only four protons away from the doubly magic $^{48}$Ca,
in $^{44}$S the self-consistent mean-field calculation predicts a coexistence of 
prolate and oblate minima separated by a rather low barrier ($ < 1$ MeV). 
For $^{42}$Si the energy surface displays a deep oblate minimum at
$(\beta,\gamma)=(0.35, 60^\circ)$, whereas a  prolate equilibrium minimum 
at $(\beta,\gamma)=(0.45, 0^\circ)$ is predicted in the very neutron-rich nucleus $^{40}$Mg.
Similar results for the quadrupole deformation energy surfaces
were also obtained in self-consistent Hartree-Fock-Bogoliubov (HFB) studies 
based on the finite-range and density-dependent Gogny interaction D1S~\cite{Delaroche10,RE.11}.

%--------------------------------------------------------------------------------------------
%--------------------------------------------------------------------------------------------

Rapid transitions between equilibrium shapes in a chain of isotones (isotopes)  
are governed by the evolution of the shell structure of single-nucleon
orbitals. In particular, one or more close-lying deformed minima can develop  
as a result of the occurrence of gaps or regions of low single-particle level density
around the Fermi surface at finite deformation. In the present analysis we illustrate 
this phenomenon with the example of shape coexistence in $^{44}$S.

There are many experimental indications  for the transitional nature of $^{44}$S:
the low-lying $0_2^+$ and $2_1^+$ excited states, the large value of the reduced 
transition probability $B(E2, 2_1^+ \to 0_1^+)$~\cite{Grevy.05,Glasmacher.97}, 
the monopole strength $\rho^2(E0, 0_2^+ \to 0_1^+)$, and the 
reduced transition probability $B(E2, 2_1^+ \to 0_2^+)$ determined in a recent 
experiment~\cite{Force.11}. Based on these results, shell model calculations and
a simple two-level mixing model have shown that $^{44}$S exhibits 
a shape coexistence between a prolate ground state and a spherical $0_2^+$ excited state.
On the other hand, spectroscopic calculations based on the self-consistent mean-field 
approach indicate a coexistence of prolate and
oblate shapes in $^{44}$S~\cite{RE.11,Li.11}.
In a very recent shell model study several inconsistencies in previous interpretations
of data for $^{44}$S have been resolved~\cite{Chevrier.14}. Using quadrupole invariants
to determine axial and triaxial shape parameters from a shell model calculation, 
a pronounced triaxial
shape for the ground state and a slightly more pronounced prolate shape for the excited
$0_2^+$ state have been predicted. It has been shown that these states display a 
large overlap in the $(\beta,\gamma)$
plane. Higher-lying members of the ground-state band show a tendency towards prolate deformation,
whereas strong fluctuations have been predicted in the band built on top of the  state $0_2^+$.
An isomeric state $4_1^+$ was recently observed in $^{44}$S~\cite{Santiago-Gonzalez.11},
and the nature of this state has been explored in a shell model study~\cite{Utsuno.15} which has
shown that this state most probably corresponds to a two-quasiparticle $K=4$ configuration.

\begin{figure}[t]
\begin{center}
\includegraphics[scale=0.6]{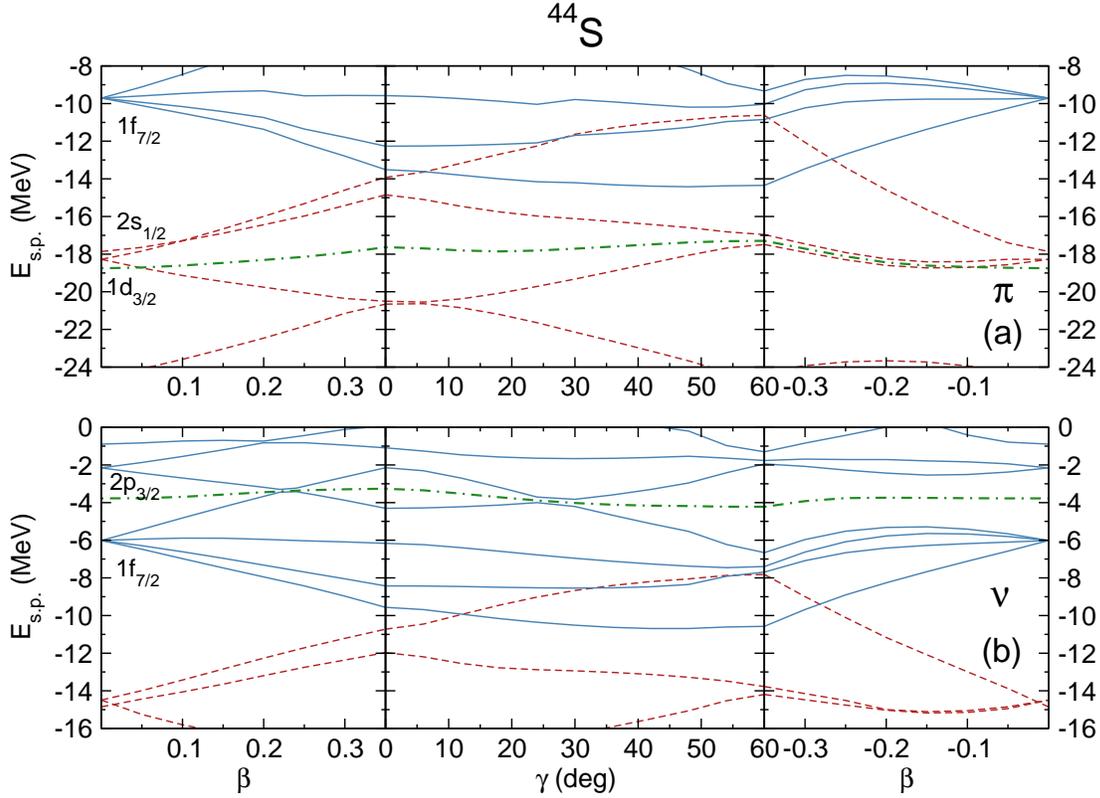}
\caption{\label{fig:s44_levels}Proton (upper panel) and neutron (lower panel) 
single-nucleon energy levels of $^{44}$S, as functions of the
deformation parameters along a closed path in the $\beta - \gamma$ plane.
Solid (blue) curves correspond to levels with negative parity, and
(red) dashed curves denote positive-parity levels.
The dot-dashed (green) curves corresponds to the Fermi levels.
The panels on the left and right display prolate ($\gamma =0^\circ$) and oblate
 ($\gamma =60^\circ$) axially-symmetric single-particle levels, respectively.
 In the middle panel the proton and neutron levels are
 plotted as functions of $\gamma$ for a fixed value 
 $|\beta|=0.35$. }
\end{center}
\end{figure}

In Fig.~\ref{fig:s44_levels} we plot the proton (upper panel) and neutron (lower panel) single-particle levels in the canonical basis for $^{44}$S. 
Solid (blue) curves correspond to levels with negative parity, and dashed (red) curves denote positive-parity levels. 
The dot-dashed (green) curves correspond to 
the Fermi levels. The proton and neutron levels are plotted as functions of the deformation parameters along a closed 
path in the $\beta$-$\gamma$ plane. The panels on the left and right display prolate 
($\gamma = 0^\circ$) and oblate ($\gamma = 60^\circ$) axially symmetric single-particle levels, respectively. 
In the middle panel of each figure the neutron and proton levels are plotted as functions of $\gamma$ 
for a fixed value of the axial deformation $|\beta| = 0.35$ which corresponds to the position of the prolate mean-field minimum 
in $^{44}$S isotope (cf. Fig.~\ref{fig:n28_pes}). Starting from the spherical configuration, we follow the single-nucleon levels on a 
path along the prolate axis up to the approximate position of the minimum (left panel), then for this fixed value of $\beta$ the path from 
$\gamma = 0^\circ$ to $\gamma=60^\circ$ (middle panel) and, finally, back to
the spherical configuration along the oblate axis (right panel). 
Axial deformations with $\gamma=60^\circ$ are denoted by negative values of $\beta$. 
This figure illustrates the principal
characteristics of structural changes in neutron-rich $N=28$ nuclei: the near degeneracy
of the $d_{3/2}$ and $s_{1/2}$ proton orbitals, and the reduction of the size of the
$N=28$ shell gap \cite{Sorlin.10}. Between the doubly magic $^{48}$Ca and $^{44}$S the 
spherical gap $N=28$ decreases from 4.73 MeV to 3.86 MeV, respectively. 
Consequently, the largest gap
between neutron states at the Fermi surface 
is located on the oblate axis (lower panel of Fig.~\ref{fig:s44_levels}), 
and we also notice the increased density of single-neutrons levels close to the
Fermi surface at $\gamma \approx 30^\circ$ which leads to formation of 
the potential barrier in the triaxial region.
For the protons (upper panel of Fig.~\ref{fig:s44_levels}), the largest gap
is located on the prolate axis. The competition between pronounced 
proton-prolate and neutron-oblate energy gaps is at the origin
of the coexistence of deformed shapes in $^{44}$S.

%--------------------------------------------------------------------------------------------------
\begin{figure}[]
\centering
\includegraphics[scale=0.65]{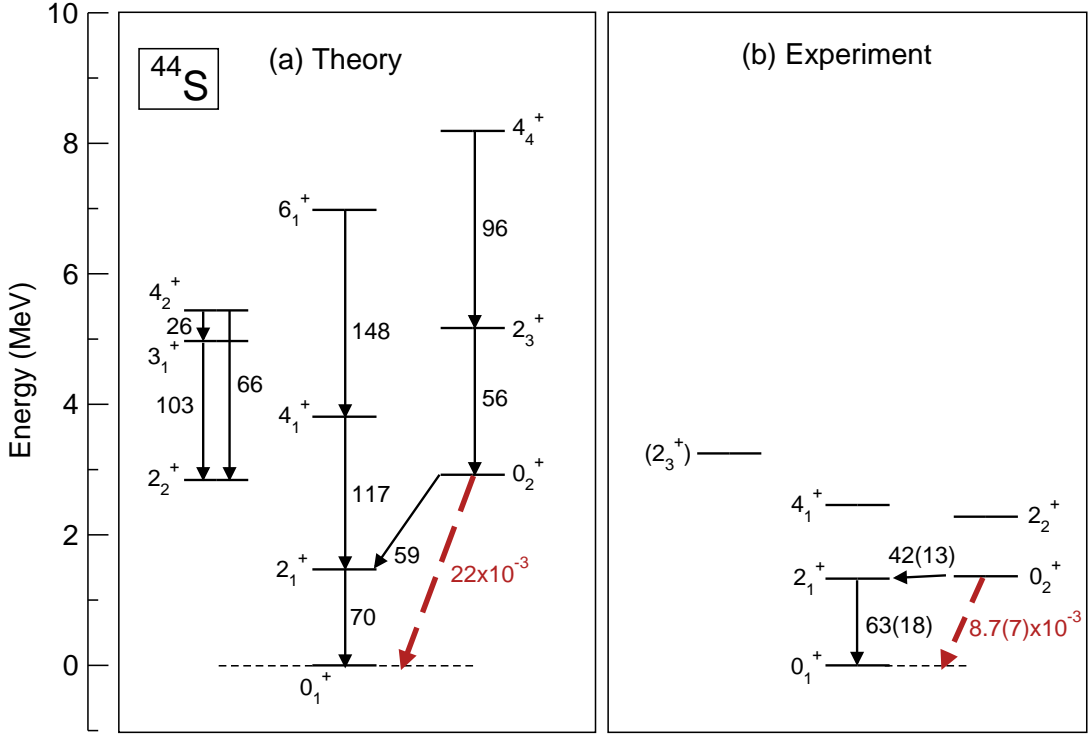}
\caption{\label{fig:s44_spec}The theoretical excitation spectrum of $^{44}$S (left), compared to 
data~\cite{Glasmacher.97,Force.11,Santiago-Gonzalez.11} (right). The
$B(E2)$ values are in units of $e^2$fm$^4$. The comparison of
the monopole transition strength $\rho^2(E0;0_2^+ \to 0_1^+)$ 
with the experimental value is also included.  }
\end{figure}
%--------------------------------------------------------------------------------------------------

Starting from constrained self-consistent solutions of the relativistic 
Hartree-Bogoliubov (RHB) equations at each point on the 
energy surfaces (Fig.~\ref{fig:n28_pes}), we calculate 
the mass parameters $B_{\beta \beta}$, $B_{\beta \gamma}$, $B_{\gamma \gamma}$,
the three moments of inertia $\mathcal{I}_k$, as well as the zero-point energy corrections,  
that determine the collective Hamiltonian~(\ref{hamiltonian-quad}).
The diagonalization of the resulting Hamiltonian yields the excitation energies and reduced transition
probabilities. Physical observables are calculated in the full configuration space and there are no 
effective charges in the model. In Fig.~\ref{fig:s44_spec} we display the low-energy spectrum of $^{44}$S 
in comparison to available data for the excitation energies, reduced electric quadrupole
transition probabilities $B(E2)$ (in units of $e^2$fm$^4$), and the electric monopole transition
strength $\rho^2(E0;0_2^+ \to 0_1^+)$. The model reproduces both the excitation energy and
the reduced transition probability $B(E2; 2_1^+ \to 0_1^+)$ for the first excited
state $2_1^+$. The theoretical value for $B(E2;0_2^+ \to 2_1^+)$ is also in good agreement with
the data. However, the calculated excitation energy of the $0_2^+$ state is much 
higher than the experimental counterpart, and the monopole
transition strength $\rho^2(E0;0_2^+ \to 0_1^+)$ overestimates the
experimental value considerably. This indicates that there is probably more mixing between the 
theoretical states $0_1^+$ and $0_2^+$ than what can be inferred from the data. We also 
note that very recently the low-lying state $4_1^+$ has been interpreted 
as a $K = 4$ isomer dominated by the two-quasiparticle configuration 
$\nu \Omega^\pi  = 1/2^-  \otimes  \nu \Omega^\pi  = 7/2^-$ \cite{Utsuno.15}, a configuration 
not included in our collective model space.

\begin{figure}[t]
\begin{center}
\begin{tabular}{ccc}
\hspace{-1cm}\includegraphics[scale=0.3]{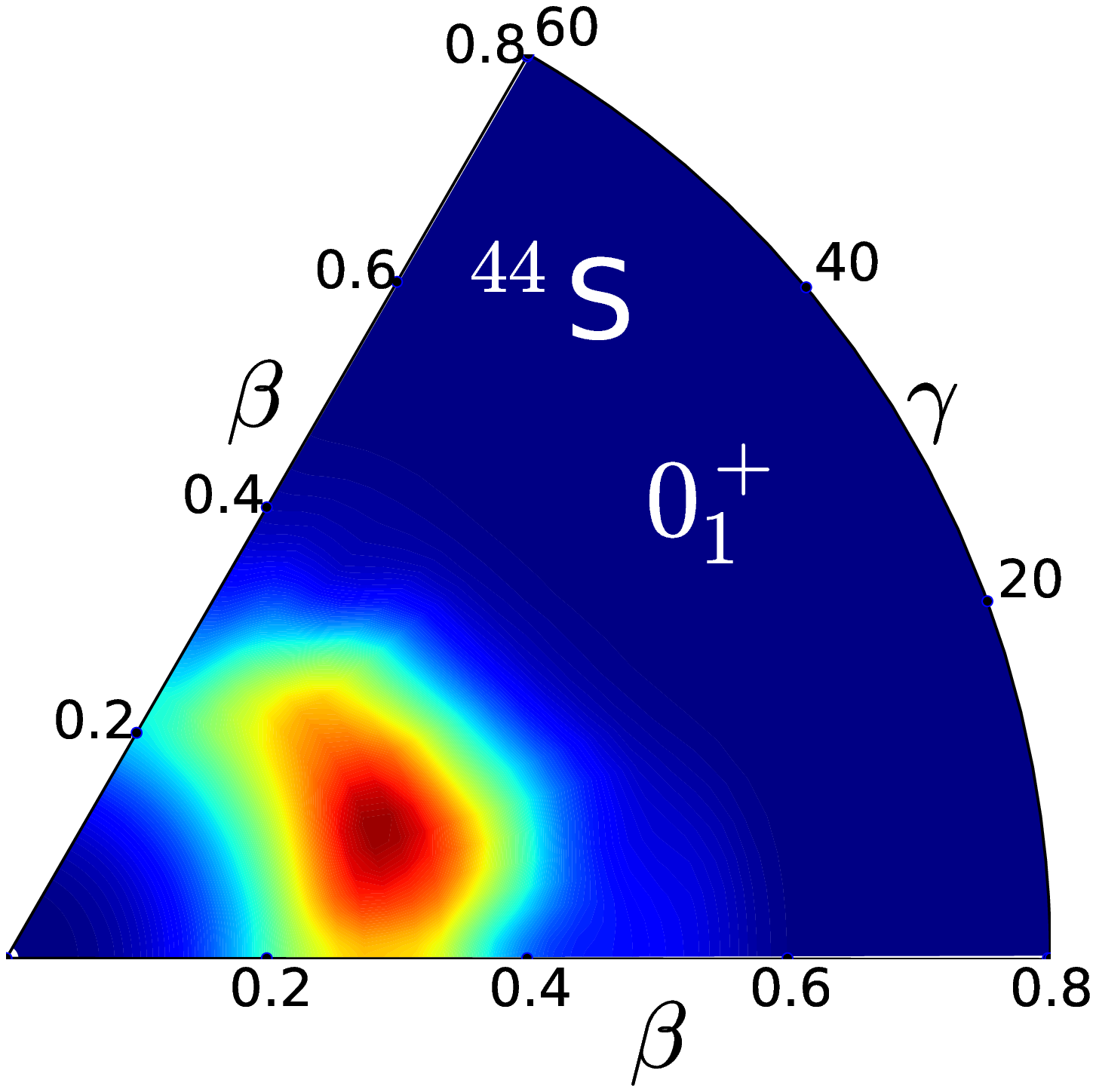}&
\hspace{-1cm}\includegraphics[scale=0.3]{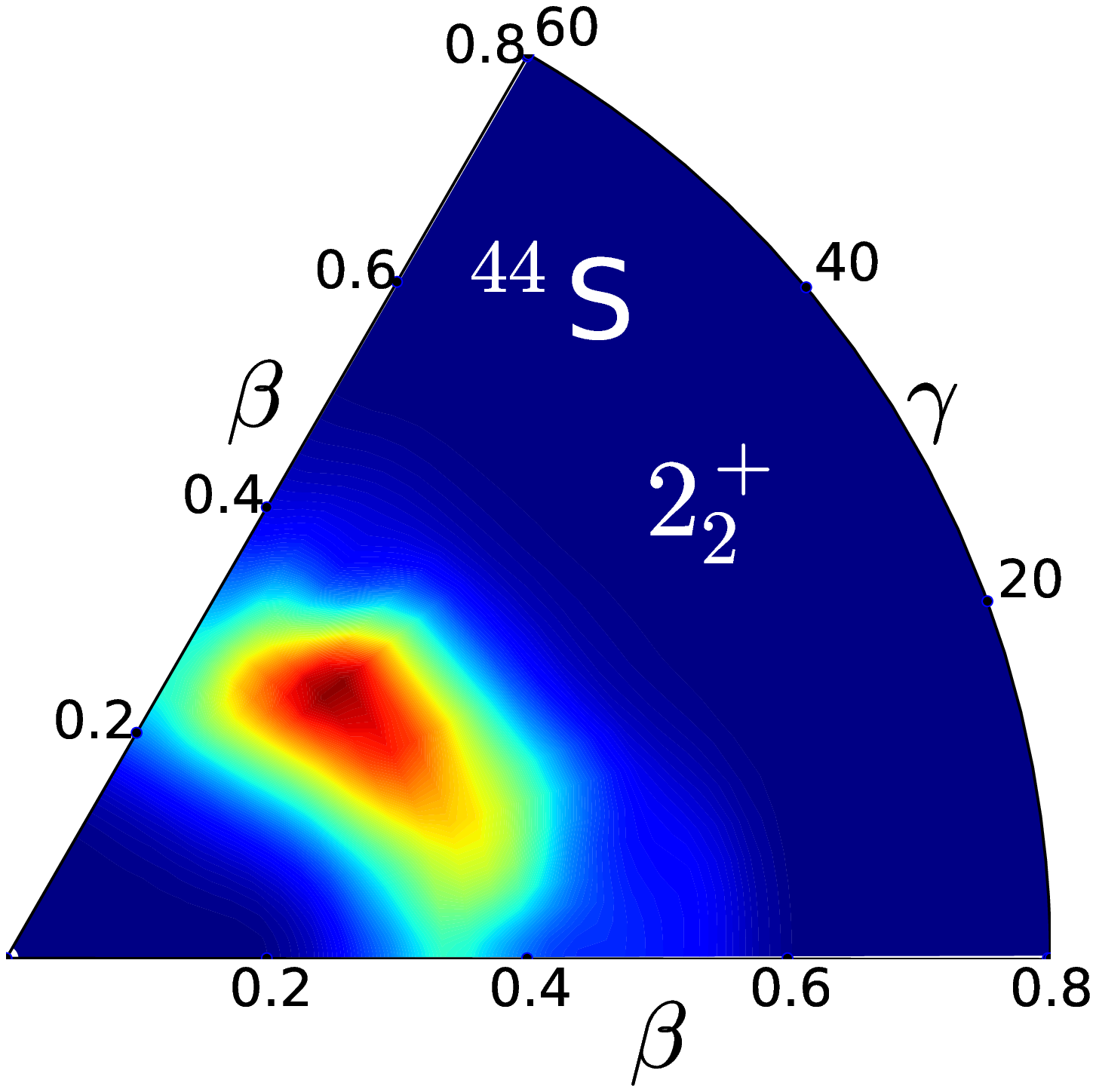}&
\hspace{-1cm}\includegraphics[scale=0.3]{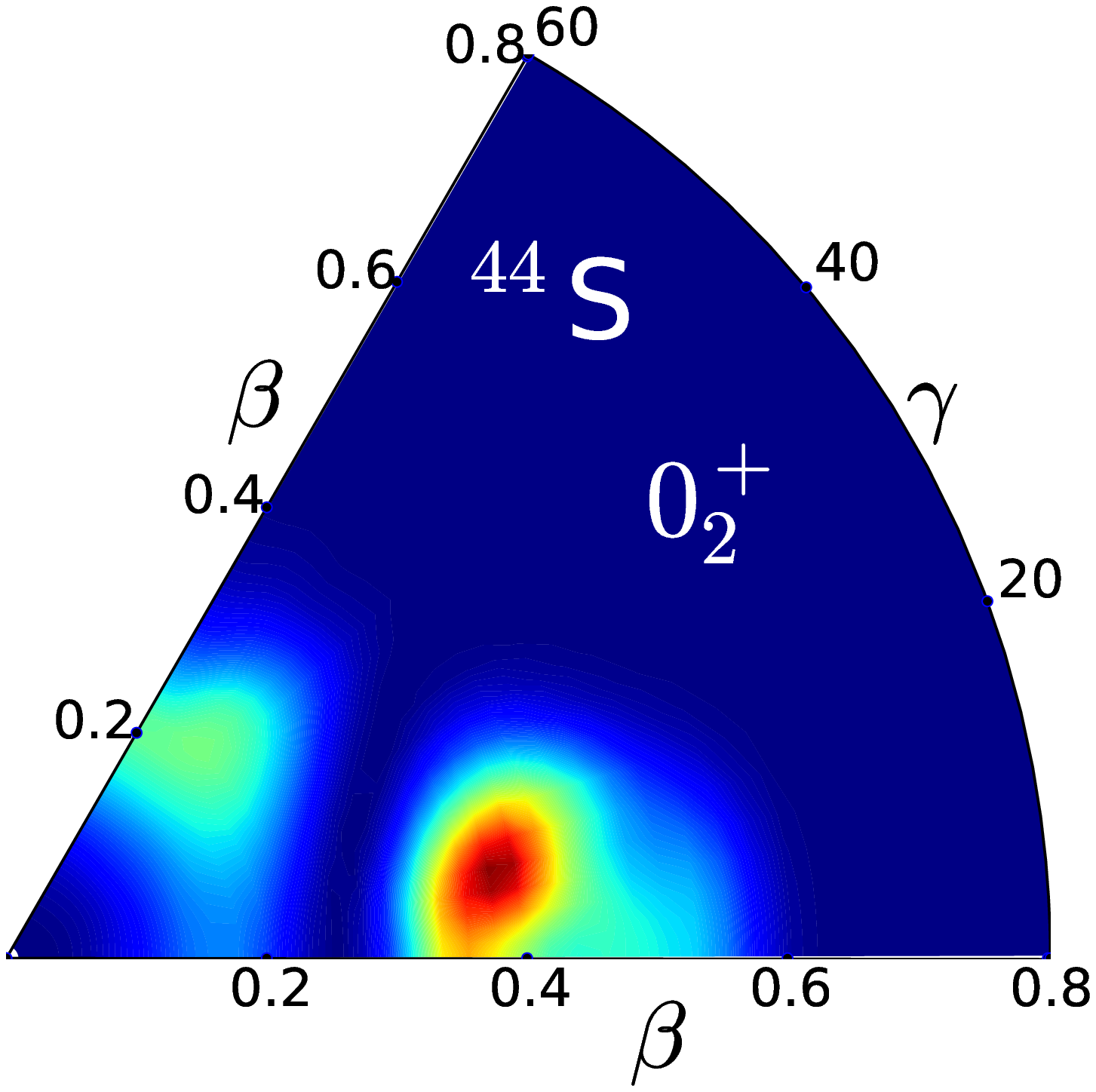} \\
\hspace{-1cm}\includegraphics[scale=0.3]{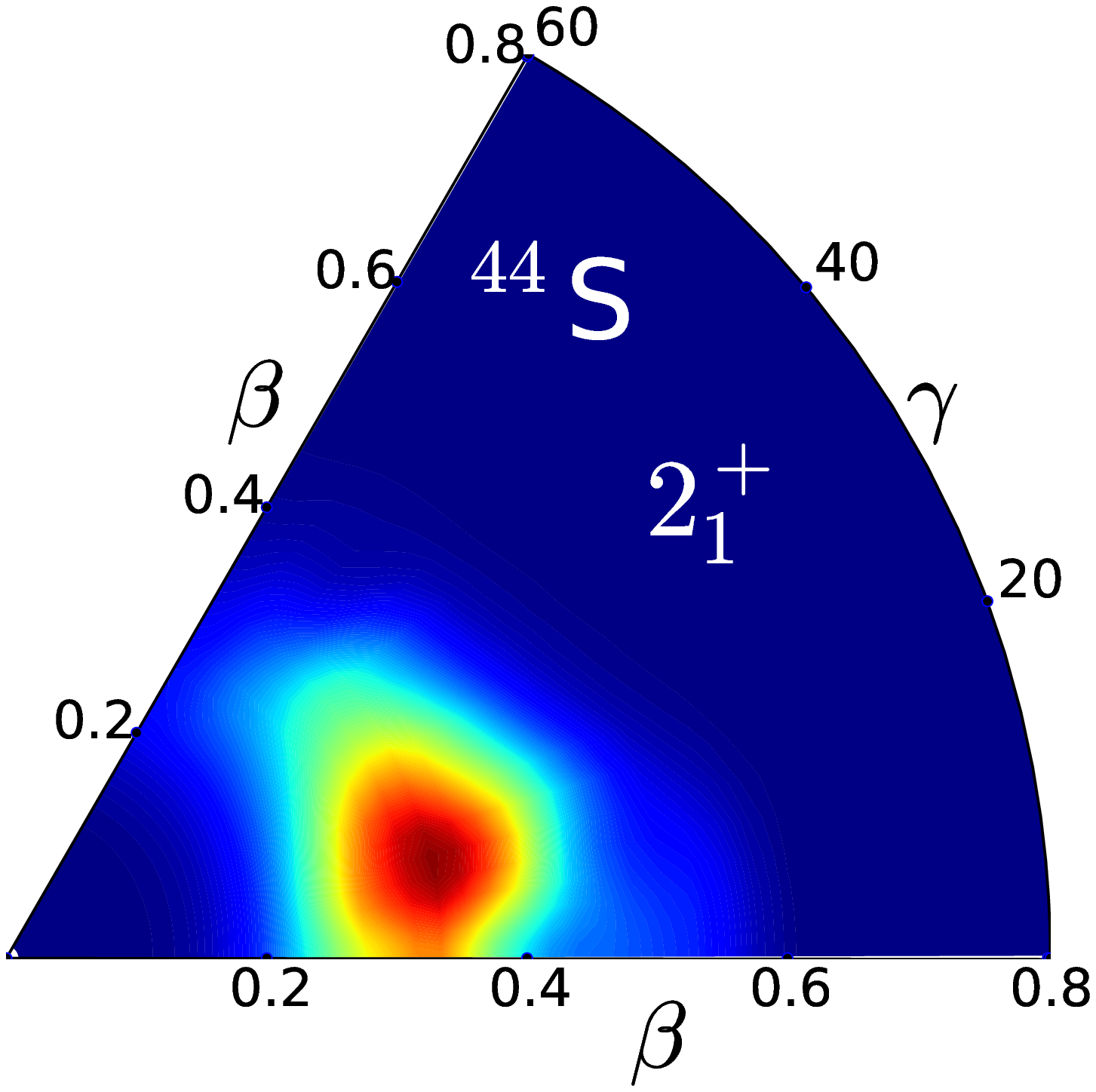}&
\hspace{-1cm}\includegraphics[scale=0.3]{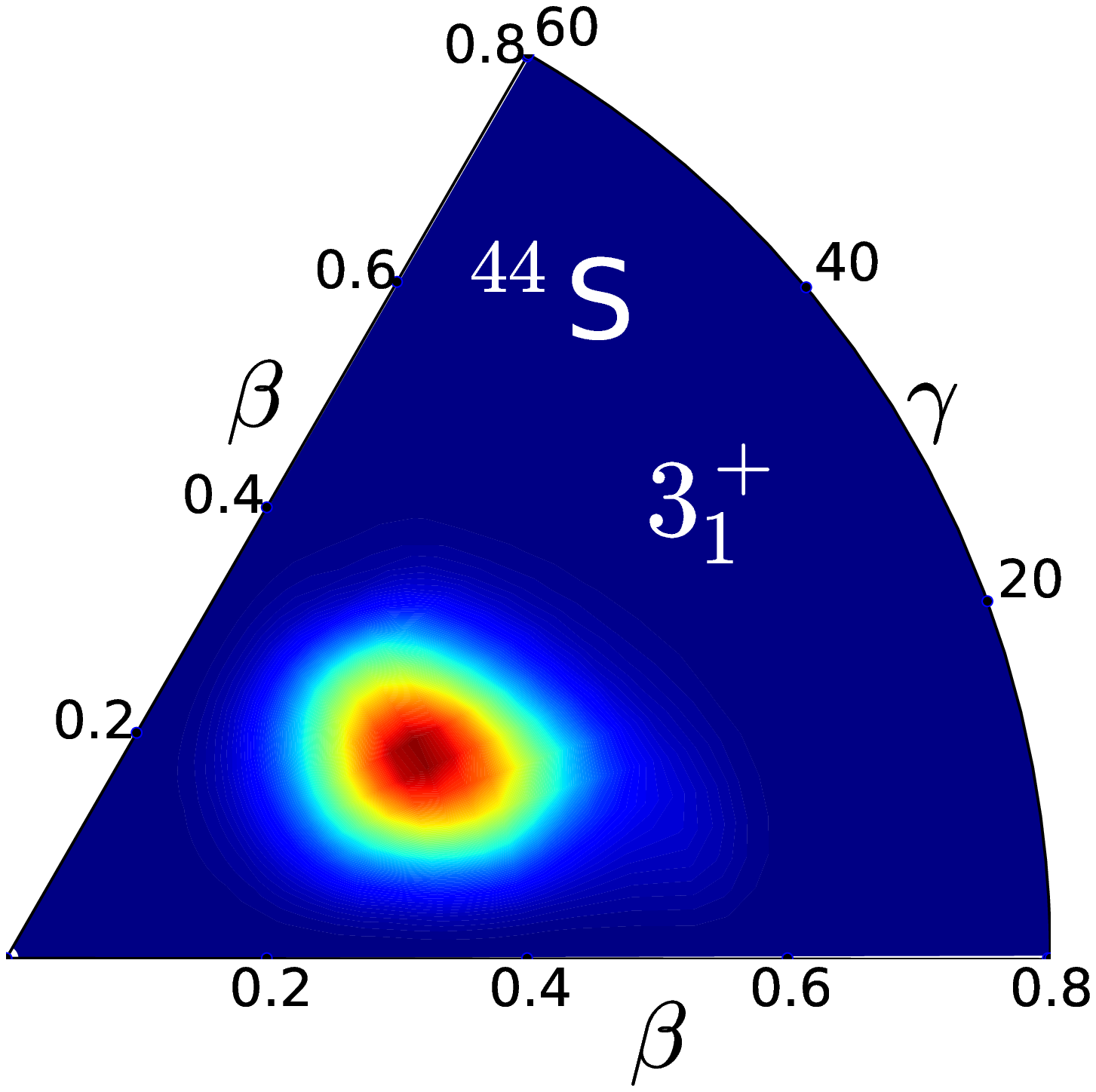}&
\hspace{-1cm}\includegraphics[scale=0.3]{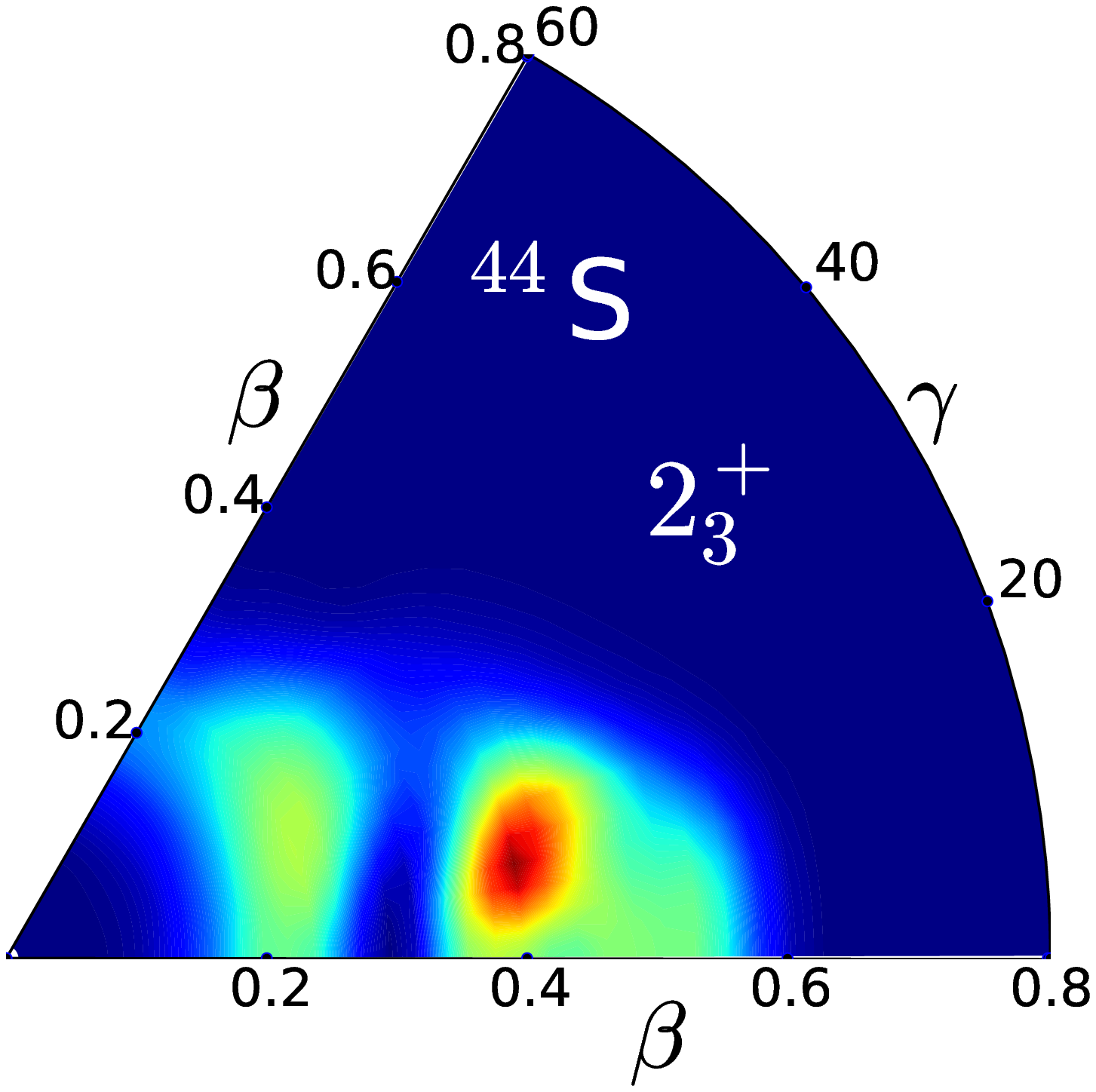} \\
\hspace{-1cm}\includegraphics[scale=0.3]{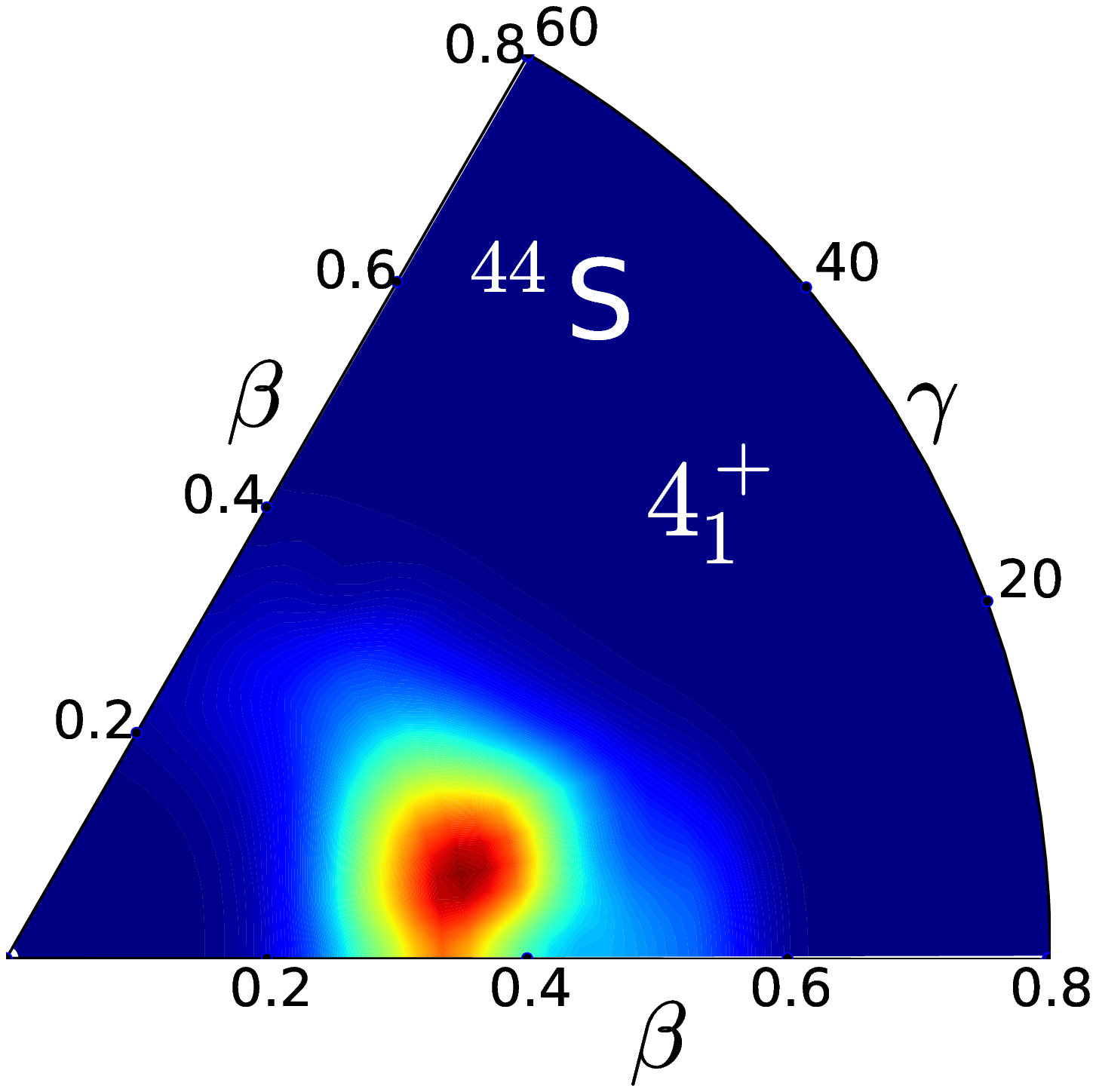}&
\hspace{-1cm}\includegraphics[scale=0.3]{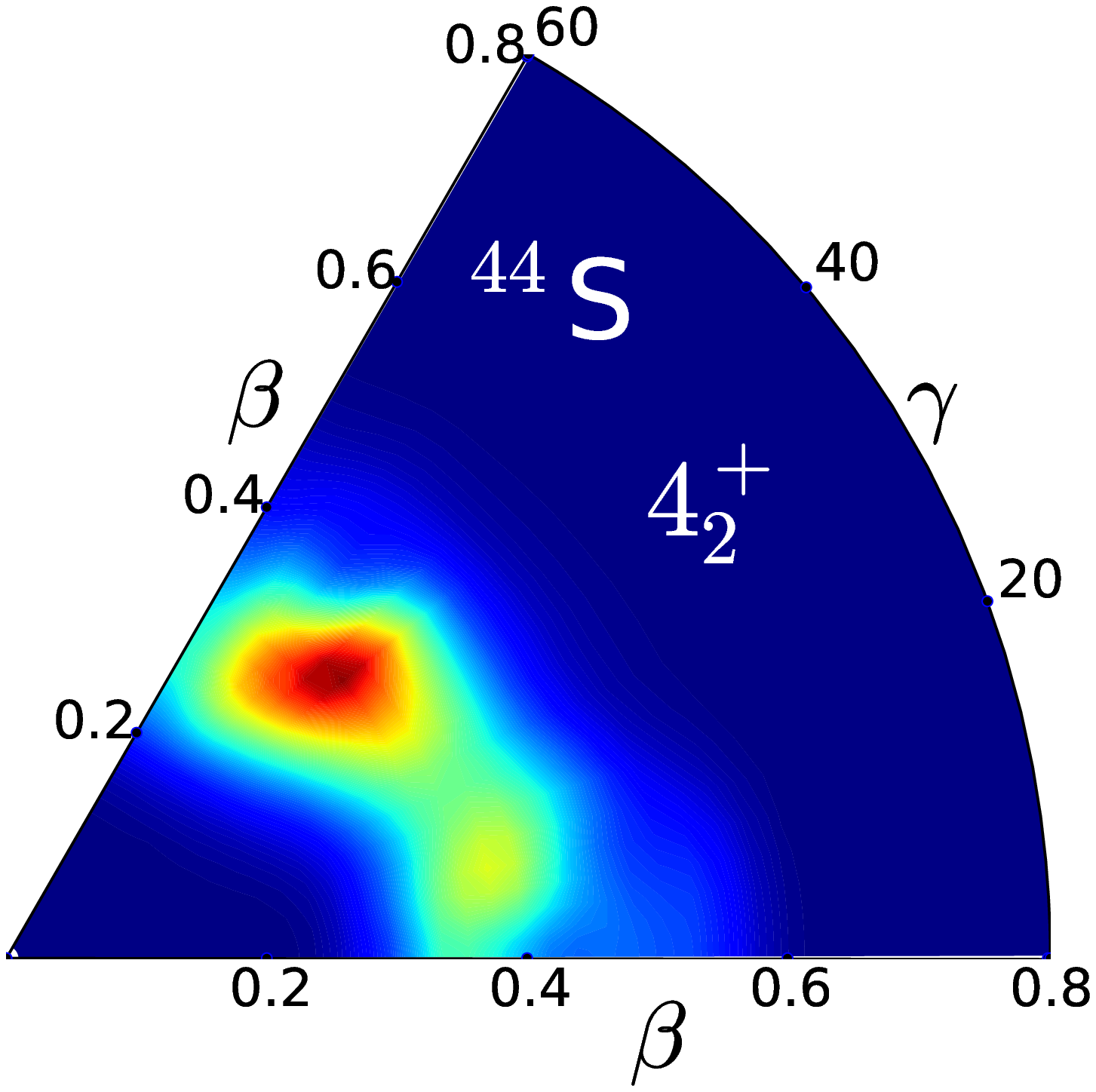}&
\hspace{-1cm}\includegraphics[scale=0.3]{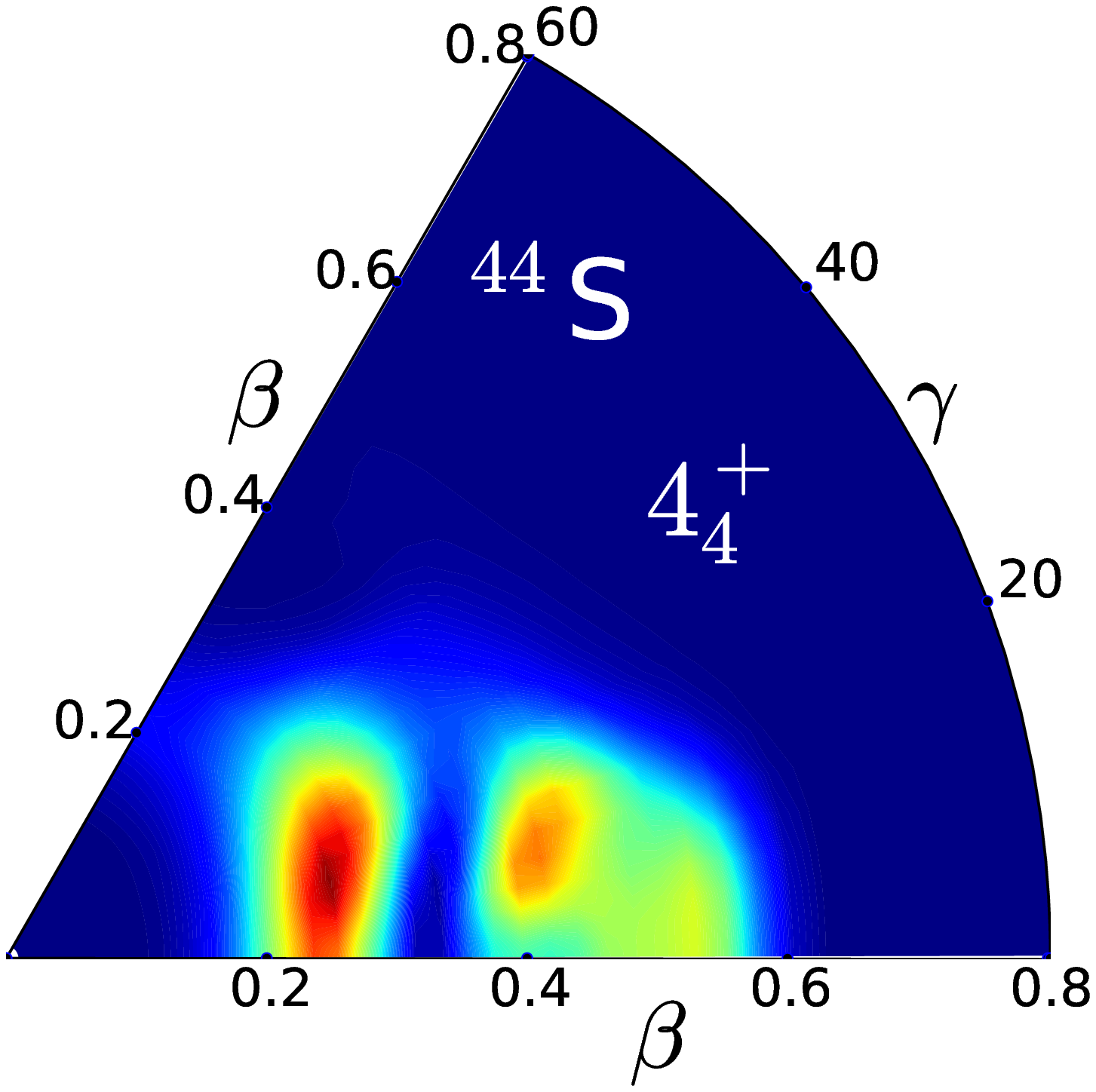}
\end{tabular}
\caption{\label{fig:s44_waves} Probability distributions Eq.~(\ref{eq:probability})
in the $\beta-\gamma$ plane for the lowest collective states of $^{44}$S.}
\end{center}
\end{figure}

To illustrate the degree of configuration mixing and shape coexistence 
in $^{44}$S, in Fig.~\ref{fig:s44_waves} we plot the
probability density distributions for the lowest three states in the ground-state band (left), 
the (quasi) $\gamma$-band (middle), and the excited band built on the state $0_2^+$ (right). 
For a given collective state, the probability distribution
in the $(\beta,\gamma)$ plane is defined as
\begin{equation}
\rho_{I\alpha}(\beta,\gamma) = \sum_{K \in \Delta I}{ \left| \psi_{\alpha K}^I(\beta,\gamma)\right|^2\beta^3 },
\label{eq:probability}
\end{equation}
with the summation over the allowed set of values of the projection $K$ of the angular 
momentum $I$ on the body-fixed symmetry axis, and 
with the normalization
\begin{equation}
\int_0^\infty{\beta d\beta \int_0^{2\pi}{ \rho_{I\alpha}(\beta,\gamma)|\sin{3\gamma}|d\gamma }}=1.
\end{equation}

\begin{table}
\caption{\label{tab:s44_K_spec}Percentage of the $K=0$ and $K=2$ components (projection of the
angular momentum on the body-fixed symmetry axis) for the collective wave functions of the
three lowest $2^+$ states in $^{44}$S, and the corresponding spectroscopic quadrupole moments
(in $e$fm$^2$).}
\centering
\begin{tabular}{cccc}
\hline\hline
 &  $K = 0$ & $K = 2$ & $Q_{spec}$ \\ \hline
 $2_1^+$ &  $89\%$ & $11\%$& $-10.8$\\
 $2_2^+$ & $21\%$ & $79\%$ & $8.2$\\
$2_3^+$  & $78\%$ & $22\%$ & $-7.3$\\
 \hline \hline
\end{tabular}
\end{table}

The probability distribution of the ground state $0_1^+$ displays a deformation $|\beta| \ge 0.3$,
extended in the $\gamma$ direction from the prolate $\gamma=0^\circ$ to the oblate
$\gamma=60^\circ$ shape. The $\gamma$ softness of the $0_1^+$ state reflects the 
ground-state mixing of configurations based on the prolate and oblate minima of the 
mean-field potential (cf. Fig.~\ref{fig:n28_pes}).
States of the ground-state band with higher angular momenta are progressively concentrated
on the prolate axis and the average $\beta$ deformation gradually increases because of 
centrifugal stretching. Although the $0_2^+$ state is predominantly prolate, one notices 
oblate admixtures and,
consequently, a relatively large overlap between the wave functions of $0_1^+$ and 
$0_2^+$. The mixing between these states leads to a pronounced level repulsion which 
is probably the cause for the too high excitation energy of the 
theoretical state $0_2^+$. 
The low-lying $0_2^+$ state at $1.365$ MeV and the monopole strength 
$\rho^2(E0;0_2^+ \to 0_1^+)=8.7(7)\times 10^{-3}$ have been regarded as signatures of
prolate-spherical shape coexistence in $^{44}$S~\cite{Force.11}. 
However, recent studies have re-examined the structure of $^{44}$S~\cite{Chevrier.14},
emphasizing the effect of the triaxial degree of freedom on the low-lying excitation 
structure. The probability distribution of 
$2_3^+$ is concentrated on the prolate axis and the transition strength $B(E2;2_3^+\to 0_2^+)$
is comparable to $B(E2;2_1^+\to 0_1^+)$. 
For the three lowest $2^+$ levels, in Table~\ref{tab:s44_K_spec} we list the percentage of the
$K=0$ and $K=2$ components in the collective wave functions, together with the spectroscopic
quadrupole moments. The wave functions of the states $2_1^+$ and $2_3^+$ are dominated by
the $K=0$ components, and the spectroscopic quadrupole moments indicate prolate configurations.
In contrast the positive quadrupole moment of $2_2^+$ state points to a predominantly oblate
configuration, while the $\approx 80\%$ contribution of the $K=2$ component in the wave function
shows that this state is the band-head of a (quasi) $\gamma$-band. One also notices the
close-lying doublet $3_1^+$ and the $4_2^+$, characteristic for a $K=2$ band in a $\gamma$-soft 
potential.

The example considered in this section and similar calculations reported recently have shown that  
the EDF approach provides an accurate microscopic interpretation of the reduction of the $N=28$
spherical energy gap in neutron-rich nuclei, and a quantitative description of the evolution
of shapes in $N=28$ isotones in terms of single-nucleon orbitals as functions of the
quadrupole deformation parameters $\beta$ and $\gamma$. In particular,  
the formation of the oblate neutron and prolate proton gaps in $^{44}$S, illustrated in Fig.~\ref{fig:s44_levels}, 
is at the origin of the predicted shape coexistence, in very good agreement with recent data.

%============================================================
%  Section 4
\section{\label{secIV}Lowest $\bm{0^+}$ excitations in $\bm{N \approx 90}$ rare-earth nuclei}
%============================================================

Rare-earth nuclei with neutron number  $N\approx 90$ present some of the best examples 
of rapid shape evolution and shape phase transitions \cite{HW.11,CJC.10,CMC.07}. Employing a 
consistent framework of structure models (GCM, quadrupole collective Hamiltonian) based on 
relativistic energy density functionals, in a series of studies \cite{NVL.07,LNV.09,Li.09} we 
analysed microscopic signatures of ground-state shape phase transitions in this 
region of the nuclear mass table. Phase transitions in equilibrium shapes of atomic nuclei correspond
to first- and second-order quantum phase transitions (QPT) between
competing ground-state phases induced by variation of a non-thermal
control parameter (number of nucleons) at zero temperature. In general, one observes 
a a gradual evolution of shapes with the number of nucleons, and these transitions 
reflect the underlying modifications of shell structure and interactions between valence nucleons. 
A phase transition, on the other hand, is characterised by a significant variation of one or more order
parameters as functions of the control parameter. Even though in systems composed
of a finite number of particles phase transitions are actually smoothed out, in many cases clear 
signatures of abrupt changes of structure properties are observed. A number of experiments 
over the last two decades, as well as many theoretical studies of deformation energy surfaces and 
also direct computation of observables related to order parameters, have shown that 
two-neutron separation energies, isotope shifts, energy gaps between the ground state and the 
excited vibrational states with zero angular momentum,
isomer shifts, and monopole transition strengths, exhibit sharp discontinuities at neutron number
$N=90$. 

%--------------------------------------------------------------------------------------------------
\begin{figure}[]
\centering
\includegraphics[scale=0.65]{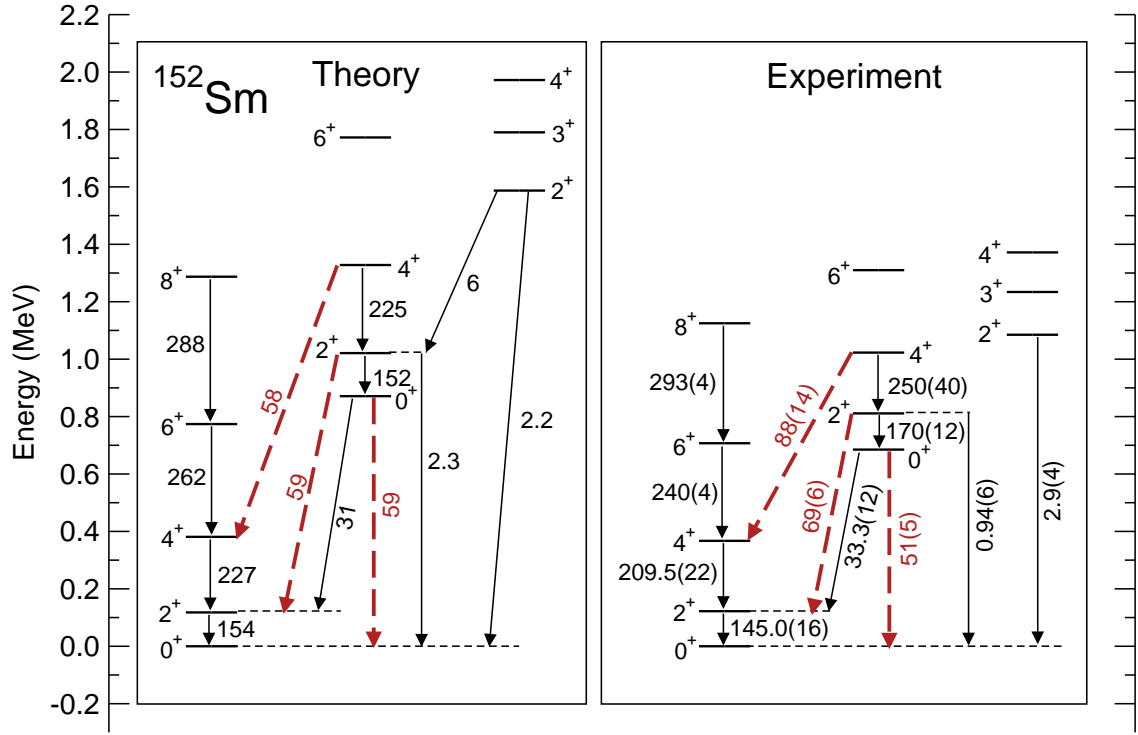}
\caption{\label{fig:sm152_spec}The theoretical excitation spectrum of $^{152}$Sm 
(left), compared to data~\cite{data}. The intraband 
and interband B(E2) values (thin solid arrows) are in Weisskopf units(W.u.), 
and (red) dashed arrows denote E0 transitions with the 
corresponding $\rho^2(E0) \times 10^3$ values.}
\end{figure}
%--------------------------------------------------------------------------------------------------
%--------------------------------------------------------------------------------------------------
\begin{figure}[]
\centering
\includegraphics[scale=0.65]{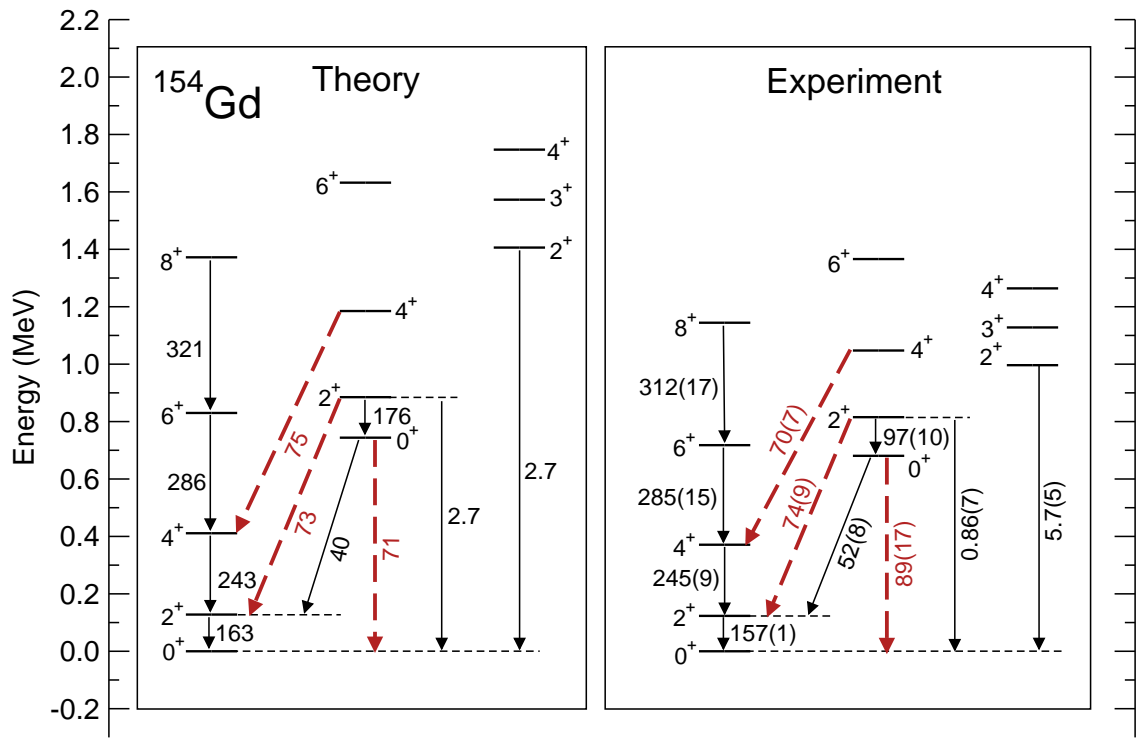}
\caption{\label{fig:gd154_spec} Same as in the caption to Fig.~\ref{fig:sm152_spec} 
but for $^{154}$Gd.}
\end{figure}
%--------------------------------------------------------------------------------------------------
%--------------------------------------------------------------------------------------------------
\begin{figure}[]
\centering
\includegraphics[scale=0.65]{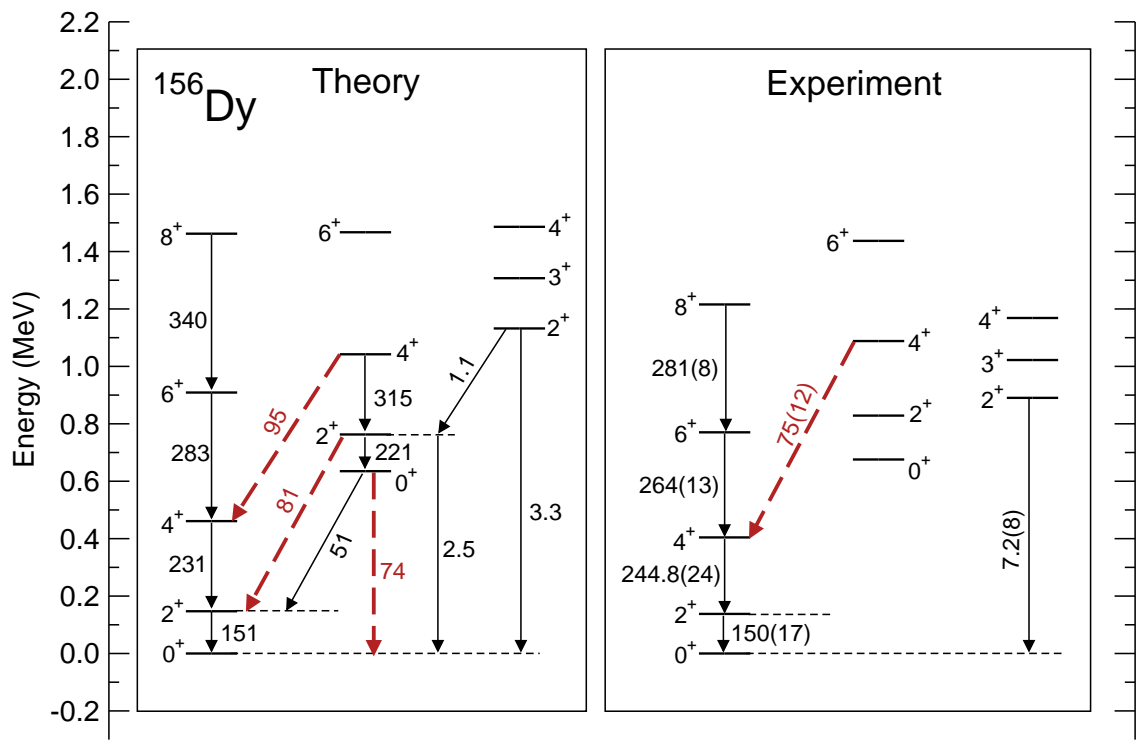}
\caption{\label{fig:dy156_spec}Same as in the caption to Fig.~\ref{fig:sm152_spec} 
but for $^{156}$Dy.}
\end{figure}
%--------------------------------------------------------------------------------------------------

In the present study we focus on the low-lying $0^+$ excitations in the deformed 
$N=90$ isotones and examine the mixing between the lowest $K=0$ bands. Traditionally the 
first excited $0^+$ level in deformed nuclei has been interpreted as a $\beta$-vibrational state, 
with the associated $K=0$ rotational $\beta$-band. However, many experimental studies 
have shown that most of the $0^+_2$ excitations are not $\beta$ vibrations. In the 
exhaustive review of properties of the lowest-lying $0^+$ states in deformed rare-earth 
nuclei \cite{Garr.01}, it was emphasized that there is no {\it a priori} reason to associate 
the $\beta$ vibration with the lowest $0^+$ excited state. A set of properties was suggested that 
the first excited $0^+$ level should exhibit in order to be labelled as a $\beta$ vibration. 
Among those: $B(E2; 0^+_\beta \to  2^+_1)$ values of $12 - 33$ W.u. or conversely 
$B(E2; 2^+_\beta \to  0^+_1)$ of $2.5 - 6$ W.u., and $\rho^2(E0;{0^+_2 \to 0^+_1}) \times 10^3$ values of $85-230$.
In our microscopic analysis of order parameters in nuclear quantum phase transitions \cite{Li.09}, 
in particular for the Nd isotopic chain, it was shown that 
the excitation energies of both $0^+_2$ and $0^+_3$ exhibit a pronounced dip at
$N=90$, which can be attributed to the softness of the potential with respect to $\beta$
deformation in $^{150}$Nd. The calculated monopole transition strengths 
exhibit a pronounced increase toward $N=90$, and the 
$\rho^2(E0;{0^+_2 \to 0^+_1})$ values remain rather large
in the deformed nuclei $^{152,154,156}$Nd, a behaviour 
characteristic for an order parameter at the point of first-order QPT.

In Figs.~\ref{fig:sm152_spec} - \ref{fig:dy156_spec} we plot the theoretical low-energy 
spectra of the $N=90$ nuclei: $^{152}$Sm, $^{154}$Gd and $^{156}$Dy, in comparison 
with data~\cite{data}. The ground-state bands, lowest $K=0$ and $K=2$ bands 
are compared to their experimental counterparts: excitation energies, intraband 
and interband $B(E2)$ values, and $E0$ transition strengths. The theoretical spectra 
comprise eigenstates of the five-dimensional quadrupole collective Hamiltonian~(\ref{hamiltonian-quad}),  
based on triaxial SCMF solutions obtained with the density functional DD-PC1 and 
a finite-range pairing force separable in momentum space. No additional parameters 
are adjusted to data, and transition rates have been calculated with bare 
charges, that is, $e_p = e$ and $e_n =0$. 

\begin{figure}[]
\begin{center}
\begin{tabular}{ccc}
\hspace{-1cm}\includegraphics[scale=0.3]{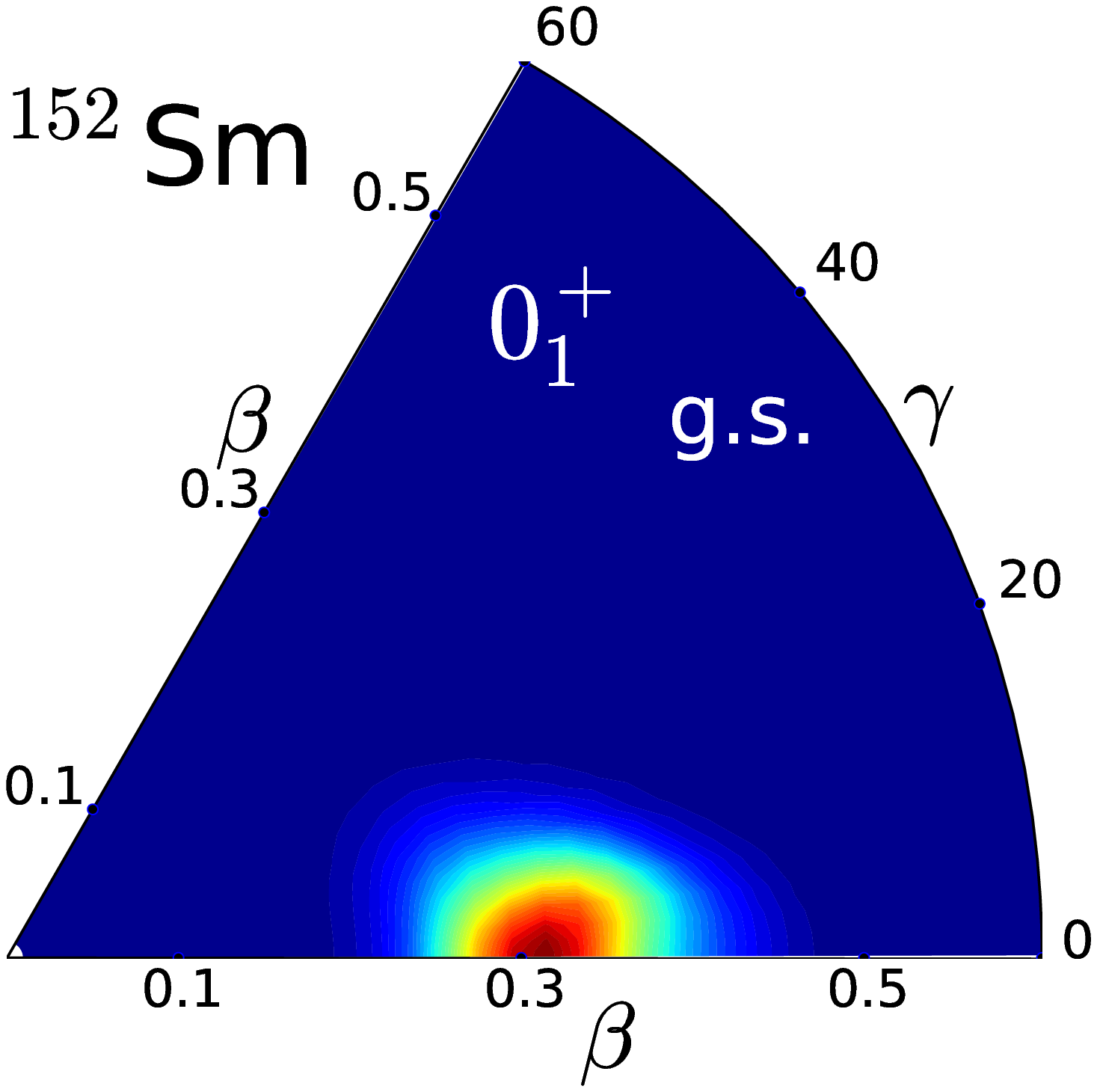}&
\hspace{-1cm}\includegraphics[scale=0.3]{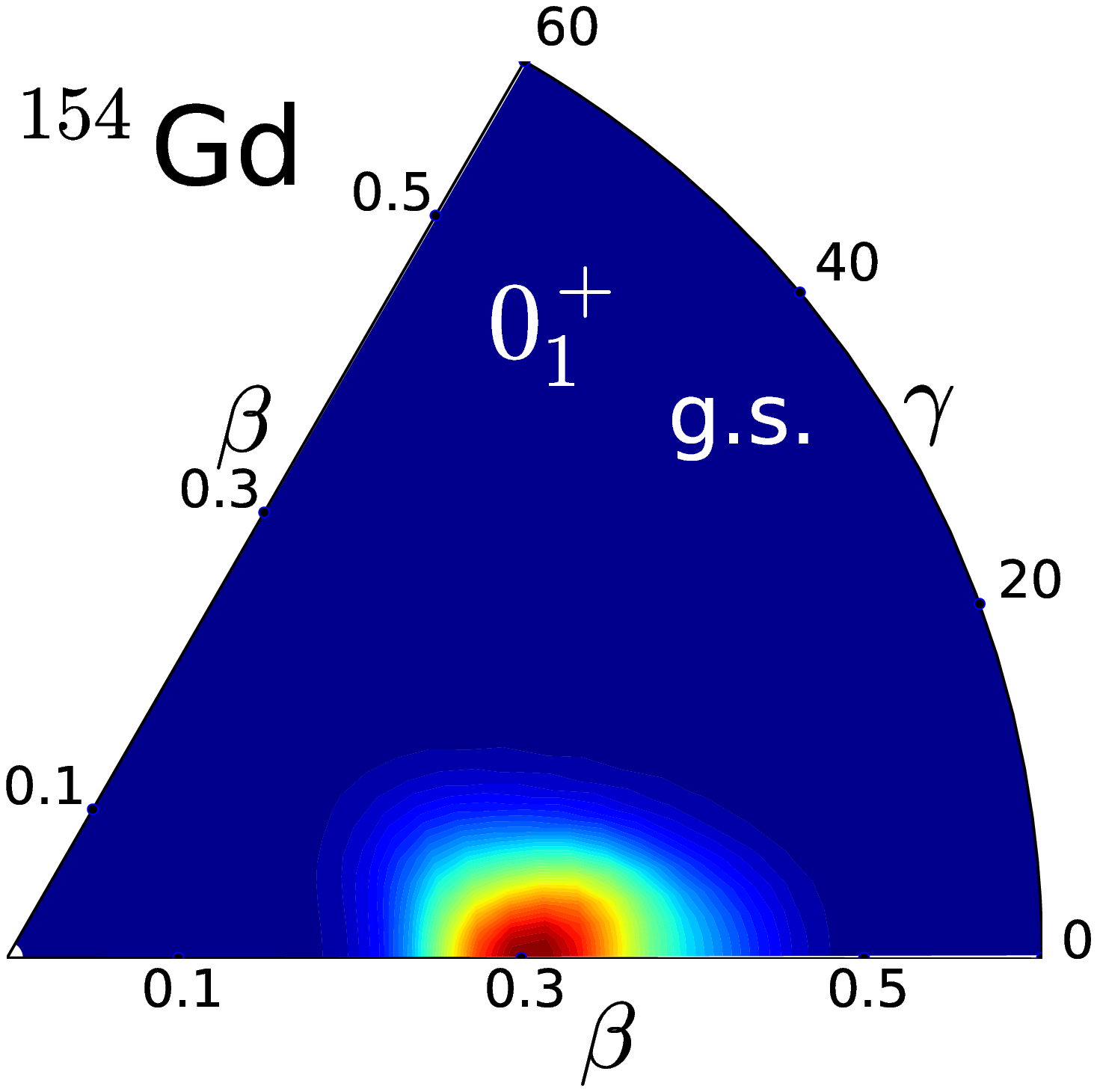}&
\hspace{-1cm}\includegraphics[scale=0.3]{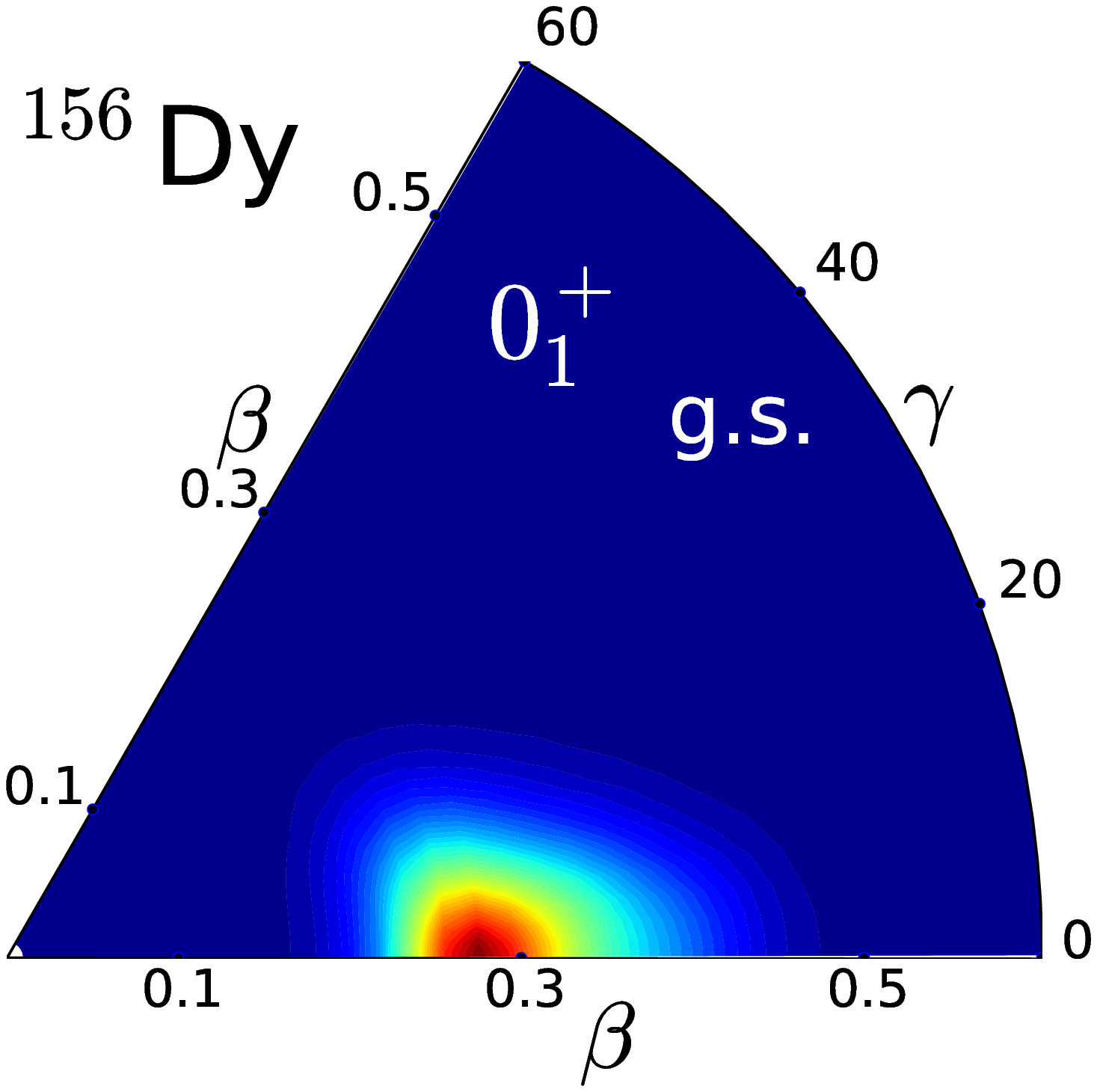}\\
\hspace{-1cm}\includegraphics[scale=0.3]{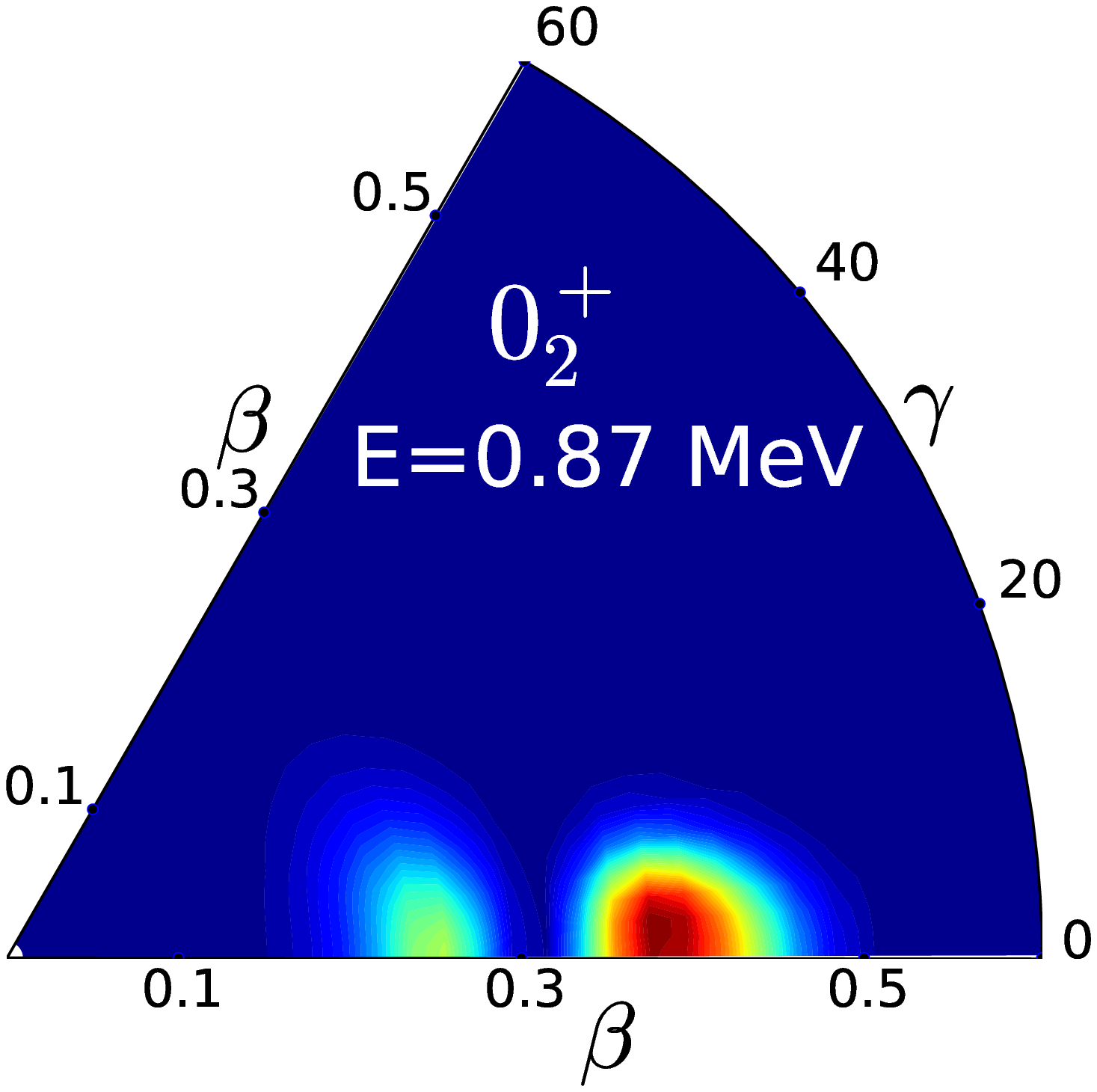}&
\hspace{-1cm}\includegraphics[scale=0.3]{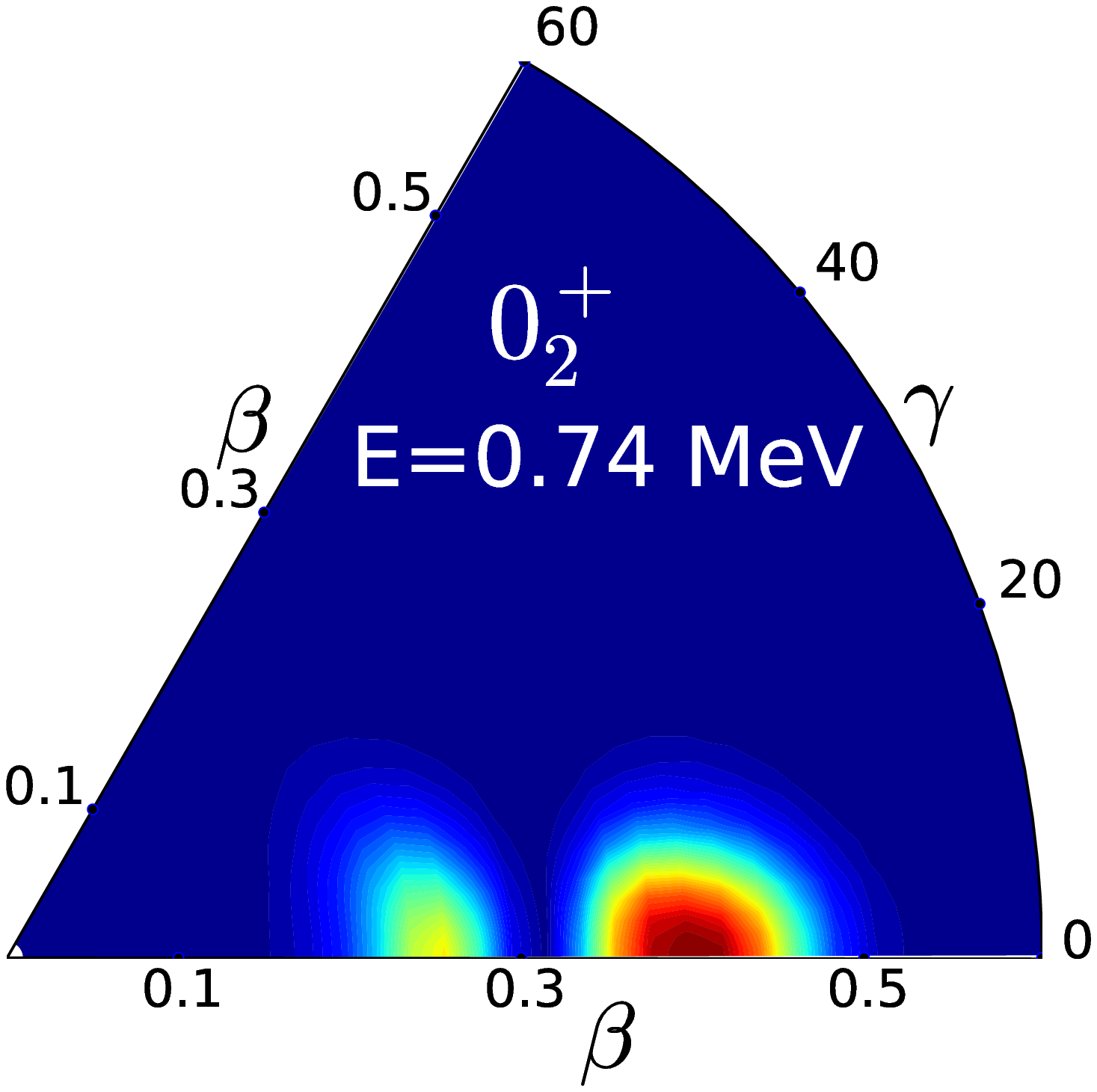}&
\hspace{-1cm}\includegraphics[scale=0.3]{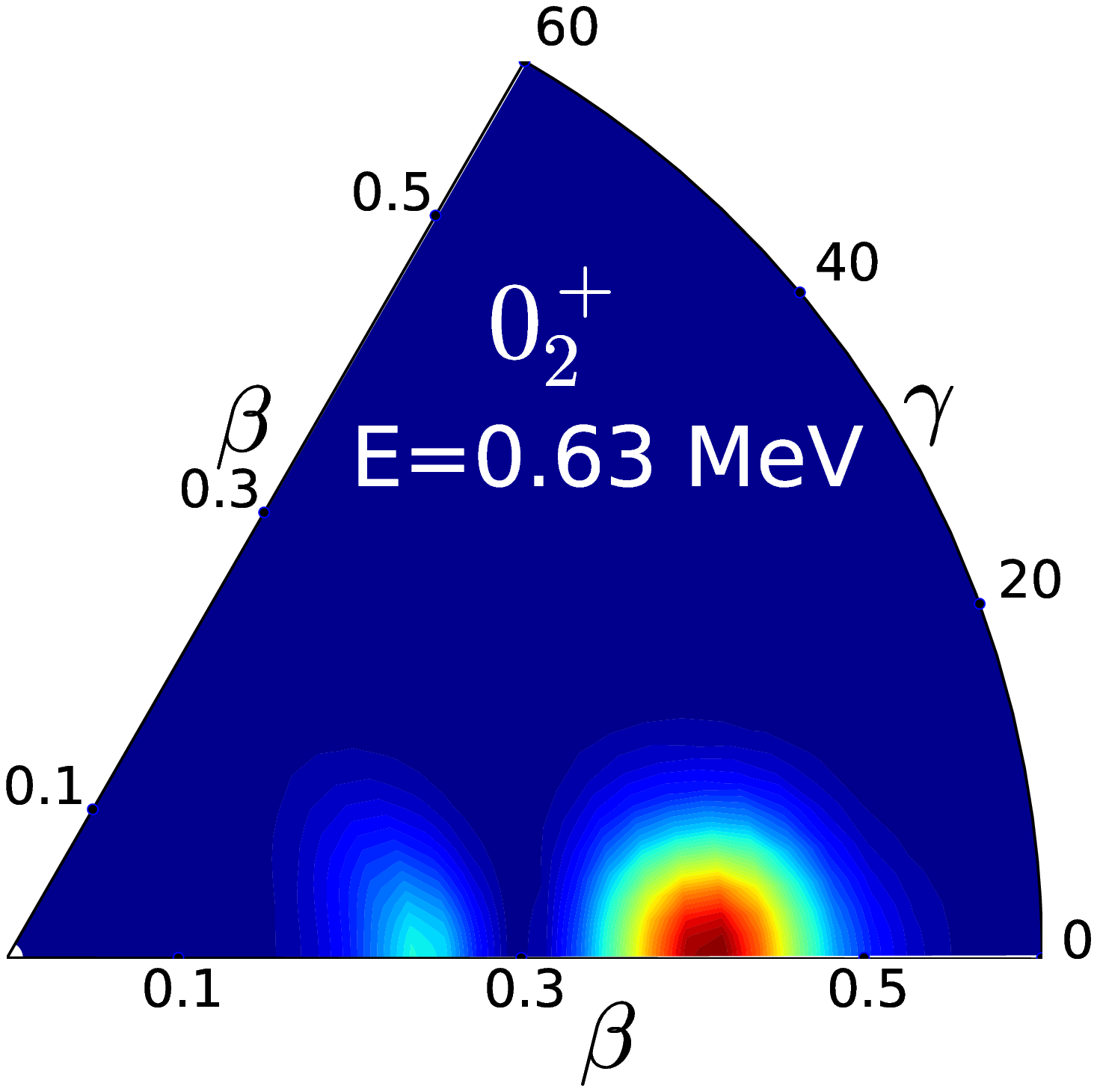}\\
\hspace{-1cm}\includegraphics[scale=0.3]{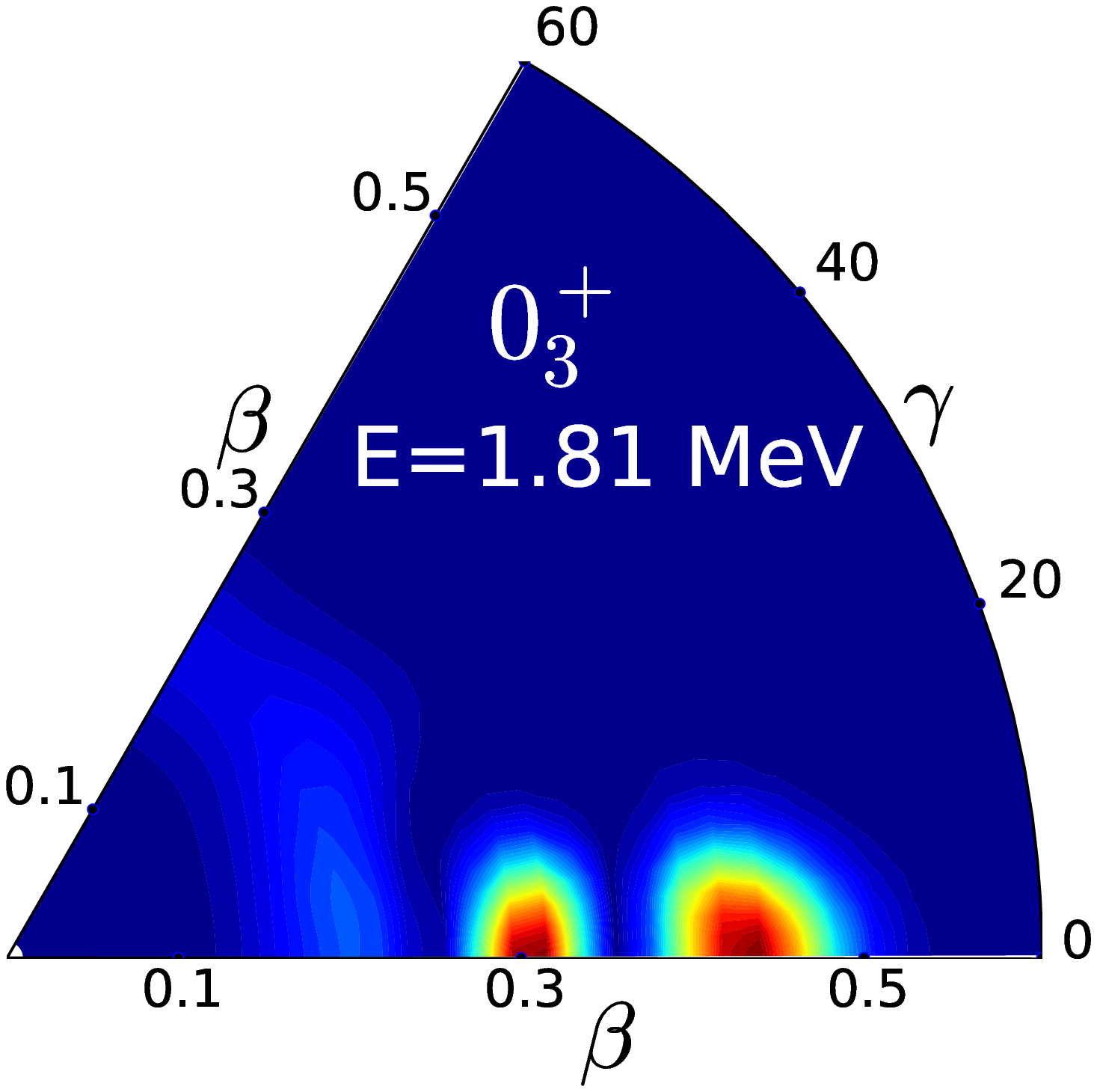}&
\hspace{-1cm}\includegraphics[scale=0.3]{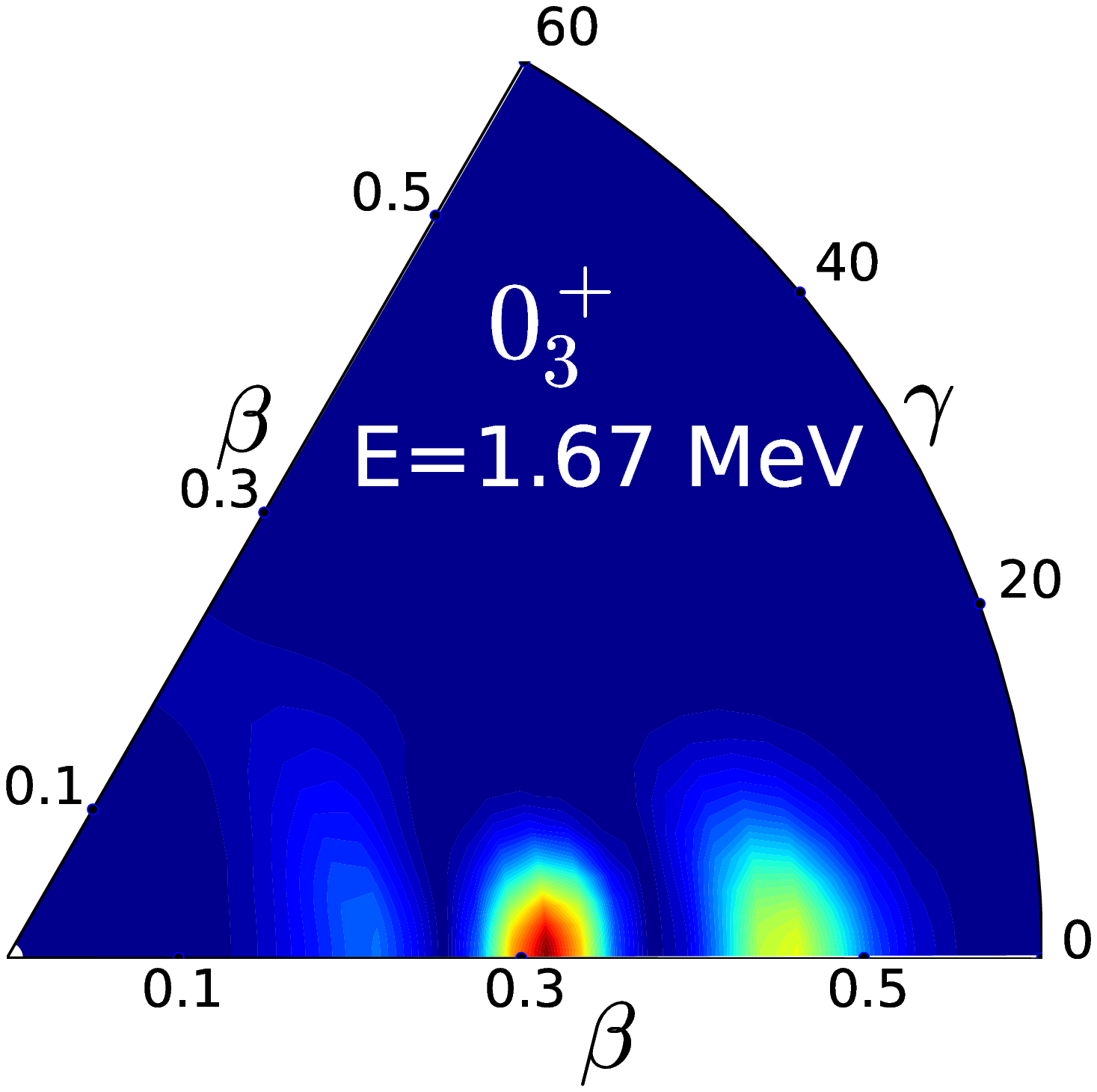}&
\hspace{-1cm}\includegraphics[scale=0.3]{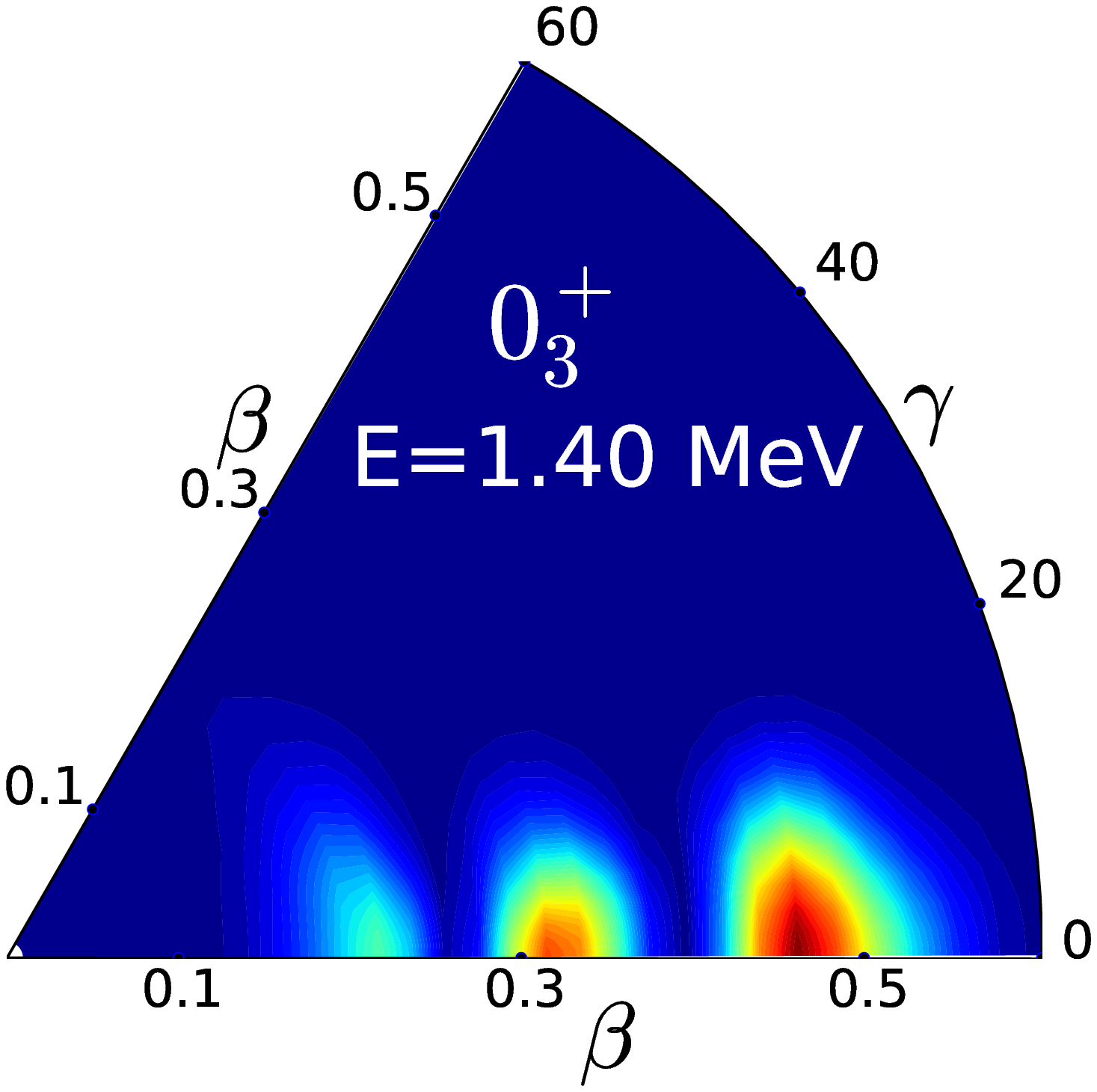}\\
\end{tabular}
\caption{\label{fig:N=90_0_states} Probability distributions Eq.~(\ref{eq:probability})
in the $\beta-\gamma$ plane for the lowest collective $0^+$ states of $^{152}$Sm, $^{154}$Gd and $^{156}$Dy.}
\end{center}
\end{figure}

\begin{figure}[t]
\begin{center}
\begin{tabular}{ccc}
\hspace{-1cm}\includegraphics[scale=0.3]{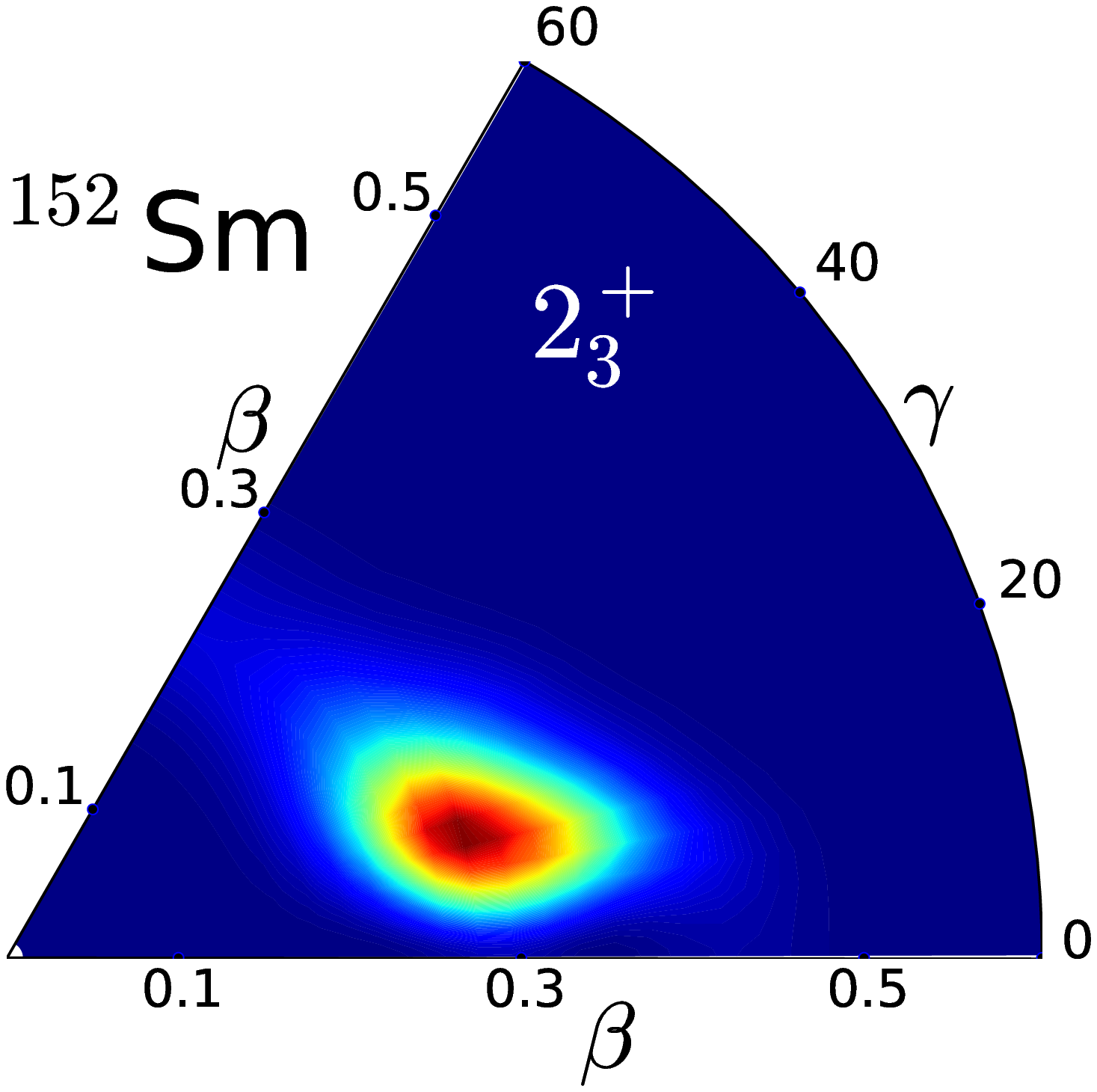}&
\hspace{-1cm}\includegraphics[scale=0.3]{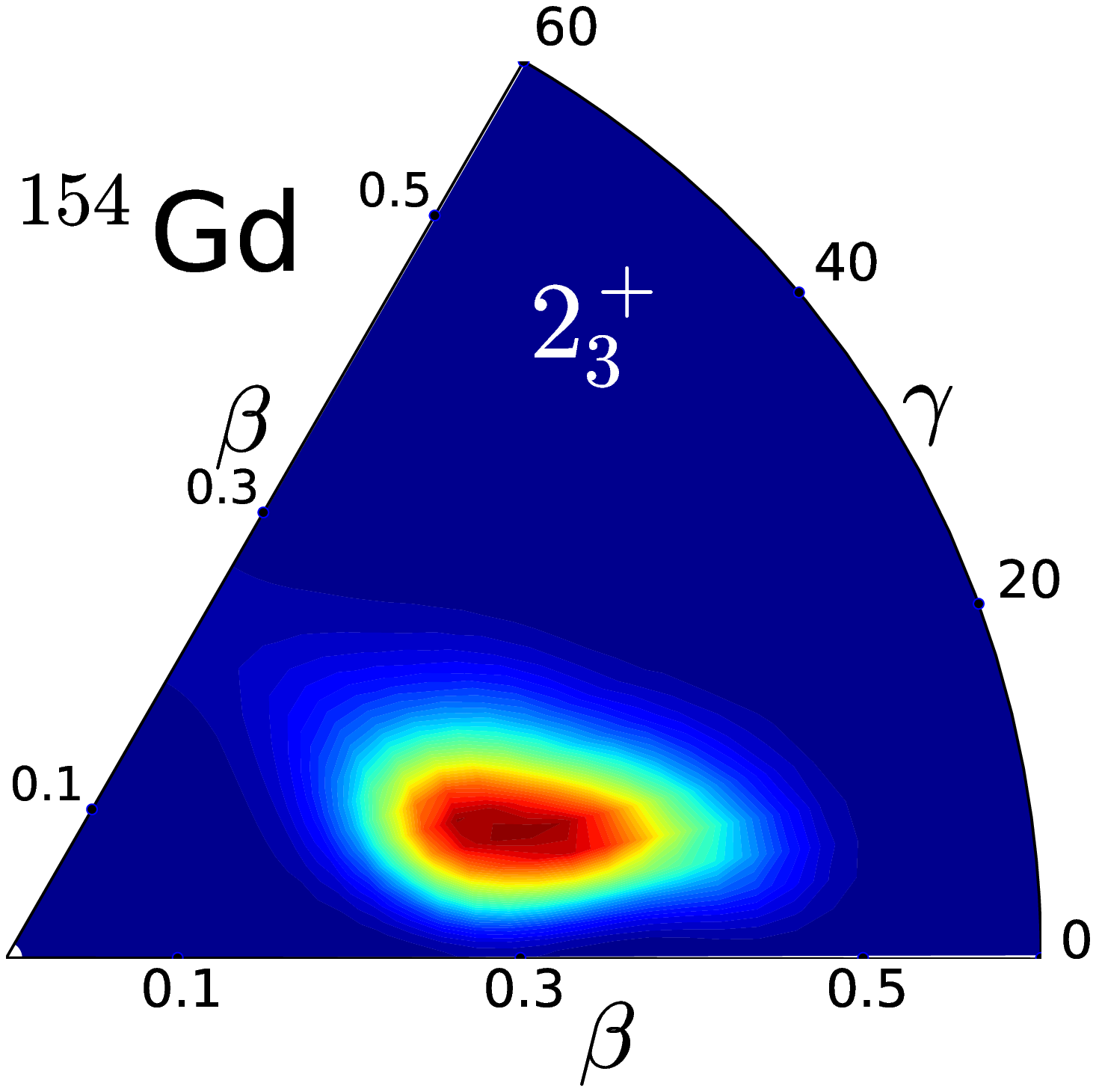}&
\hspace{-1cm}\includegraphics[scale=0.3]{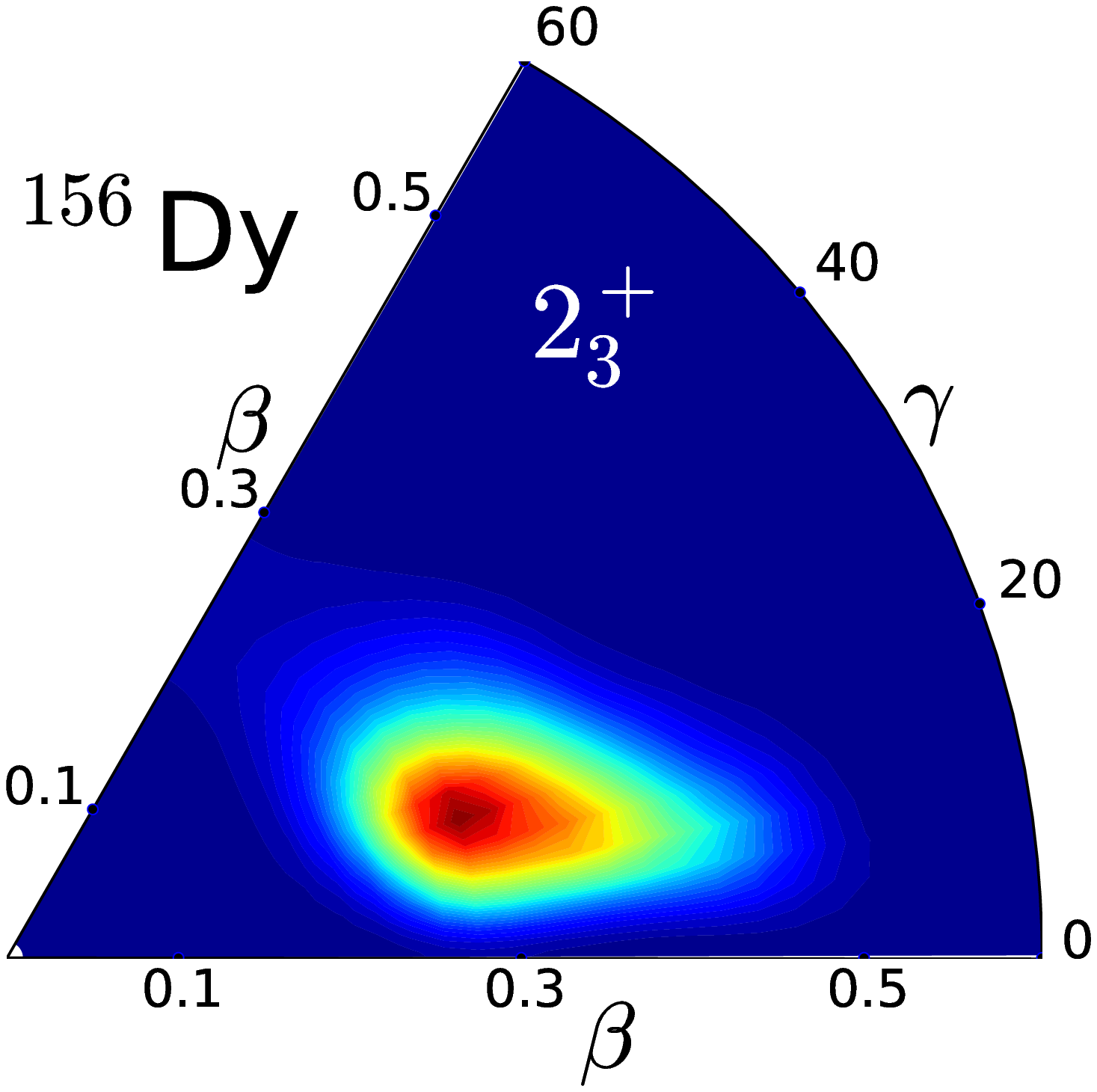}\\
\end{tabular}
\caption{\label{fig:N=90_2_3_states} Probability distributions Eq.~(\ref{eq:probability})
in the $\beta-\gamma$ plane for the band-heads of the $K=2$ $\gamma$-bands 
of $^{152}$Sm, $^{154}$Gd and $^{156}$Dy.}
\end{center}
\end{figure}

For the ground-state bands the theoretical excitation energies and $B(E2)$ values for 
transitions within the band are in very good agreement with data, except for the fact that the 
empirical moments of inertia are systematically larger than those calculated with the
collective Hamiltonian. This is a well known effect of using the simple Inglis-Belyaev 
approximation for the moments of inertia, and is also reflected in the excitation energies 
of the excited $K=0$ and $K=2$ bands \cite{DGL.10}. The wave functions, however, are not 
affected by this approximation and we note that the model 
reproduces both the intraband and interband $E2$ 
transition probabilities. The $K=2$ $\gamma$-bands are predicted at somewhat  
higher excitation energies compared to their experimental counterparts, and this is most 
probably due to the potential energy surfaces being too stiff in $\gamma$.  
The deformed rare-earth $N=90$ isotones are characterised by very low $K=0$ bands 
based on the $0^+_2$ states. In $^{152}$Sm, for instance, this state is found 
at 685 keV excitation energy, considerably below the $K=2$ $\gamma$-band. Nevertheless, 
this state has been interpreted as the band-head of the $\beta$-band \cite{Garr.01,DGL.10}. 

In $^{152}$Sm the excited $K=0$ band is calculated at moderately higher energy 
compared to data, while the agreement with experiment is very good for $^{154}$Gd and $^{156}$Dy. 
We note that a very similar excitation spectrum for $^{152}$Sm was also obtained 
with the collective Hamitonian based on the D1S Gogny interaction \cite{DGL.10}. 
Particularly important for the present study are the $E0$ transitions between the two lowest 
$K=0$ bands, and the $B(E2; 0^+_2 \to  2^+_1)$ and $B(E2; 2^+_2 \to  0^+_1)$ values. 
The available data are very accurately reproduced by the calculation and, in particular 
for $^{152}$Sm, the $E0$ transition strengths and $B(E2)$ values seem to match the 
criteria for a $\beta$-vibrational state \cite{Garr.01}.  

The $E0$ transitions strengths reflect the degree of mixing between the two lowest $K=0$ bands, 
and Figs.~\ref{fig:sm152_spec} - \ref{fig:dy156_spec} show that the theoretical values that 
correspond to transitions between the eigenstates of the collective Hamiltonian reproduce 
the empirical $\rho^2(E0)$ values. The structure of the low-lying $0^+$ states is 
analysed in Fig.~\ref{fig:N=90_0_states}, in which the probability density distributions 
are plotted in the $\beta-\gamma$ plane for the three lowest collective 
$0^+$ states of $^{152}$Sm, $^{154}$Gd and $^{156}$Dy. We note that the probability 
distributions for these states are concentrated on the prolate axis $\gamma = 0^\circ$, 
in contrast to the band-heads of the $K=2$ $\gamma$-bands, for which the probability 
density distributions are shown in Fig.~\ref{fig:N=90_2_3_states}. The dynamical  
$\gamma$-deformations of the latter clearly point to the $\gamma$-vibrational nature 
of these states. The average values of the deformation parameter 
$\beta$ for the collective ground-state wave functions of $^{152}$Sm: $<\beta> = 0.32$, 
$^{154}$Gd: $<\beta> = 0.31$, and $^{156}$Dy: $<\beta> = 0.30$, correspond to the 
minimum of the respective deformation energy surface. The corresponding values for the first 
excited $0^+$ states are: $<\beta> = 0.33$ for $^{152}$Sm, $<\beta> = 0.34$ for 
$^{154}$Gd, and $<\beta> = 0.37$ for $^{156}$Dy. For a pure harmonic $\beta$-vibrational 
state one expects that the average deformation is the same as for the ground-state, that 
the ratio of $\Delta \beta$ values for $0^+_\beta$ with respect to $0^+_1$ is $\sqrt{3}$, and that 
the probability density distribution displays one node at $<\beta>_{\rm g.s.}$ 
and two peaks of the same amplitude. The ratio of $\Delta \beta$ values for $0^+_2$ with respect to 
$0^+_1$ is: 1.6 for $^{152}$Sm, 1.53 for $^{154}$Gd, and 1.42 for $^{156}$Dy. Considering 
all these quantities, it appears that the best candidate for the $\beta$-vibrational state is 
$0^+_2$ in $^{152}$Sm. However, even in this case the probability distribution does not 
display two peaks of equal amplitude, and the shift to larger deformation is more 
pronounced in $^{154}$Gd and $^{156}$Dy. Note that for the latter two nuclei the calculated 
excitation energy of the $0^+_2$ states are in even better agreement with data. The 
probability distributions for the $0^+_3$ levels, plotted in the third row of 
Fig.~\ref{fig:N=90_0_states}, indicate the development of a second node and third peak, 
that is, the appearance of two-phonon states. One notices, however, the mixing 
with states based on $\gamma$ vibrations, which becomes even more pronounced 
for higher lying $0^+$ states, not shown in the figure. An experimental exploration 
of a possible occurrence of multiple (two) phonon intrinsic collective excitations 
in $^{152}$Sm did not find evidence for two-phonon $K^\pi = 0^+$ quadrupole 
vibrations \cite{Kulp.08}. In fact, it has been argued that an emerging pattern of repeating 
excitations built on $0^+_2$, similar to those based on the ground state, 
shows that $^{152}$Sm is an example of shape coexistence \cite{Garr.09}.

The simple analysis presented in this section illustrates the complex structure of 
excited $0^+$ levels in deformed nuclei, and the difficulties in classifying these 
states as simple collective vibrational states, that is, as one and two-phonon 
$\beta$ vibrations. A more quantitative theoretical investigation should involve 
additional effects not included in our collective Hamiltonian model, such as 
are the coupling between shape oscillations and pairing vibrations \cite{PP.93} and, 
in general, the coupling between collective and intrinsic two-quasiparticle 
excitations \cite{Ber.11}, that can lower the collective energy levels 
and improve the agreement with data \cite{data,Garr.01}. In particular, a more 
advanced model that includes coupling between collective and intrinsic two-quasiparticle 
excitations can be used to analyse excited rotational bands 
based on pairing isomers, such as those identified in 
$^{154}$Gd \cite{Kulp.03} and $^{152}$Sm \cite{Kulp.05}.

%============================================================
%  Section 5
\section{\label{secV}Quadrupole and octupole shape transition in Thorium}
%============================================================

Most deformed medium-heavy and heavy nuclei display quadrupole equilibrium shapes, 
but there are also regions of the mass table 
in which octupole deformations (reflection-asymmetric, pear-like shapes) occur.
Reflection-asymmetric shapes are characterized by the presence 
of negative-parity bands, and by pronounced electric dipole and octupole
transitions. In the case of static octupole deformations, for instance, the lowest 
positive-parity even-spin states and the negative-parity odd-spin states form an
alternating-parity band, with states connected by the enhanced
E1 transitions. In a simple microscopic picture strong octupole correlations 
arise through a coupling of orbitals near the Fermi surface with quantum numbers 
($l$, $j$) and ($l+3$, $j+3$). This leads 
to reflection-asymmetric intrinsic shapes that develop either dynamically 
(octupole vibrations) or as static octupole equilibrium deformations \cite{BN.96,BW.15}.
For example, in the case of $N\approx 134$ and $Z\approx 88$
nuclei in the region of light actinides, the coupling of the neutron orbitals based on 
$g_{9/2}$ and $j_{15/2}$, and that of the proton single-particle
states arising from $f_{7/2}$ and $i_{13/2}$, can give rise to octupole mean-field deformations. 

An interesting phenomenon are simultaneous quadrupole and octupole shape transitions. 
In a series of recent studies \cite{Nom.13,Li.13,Nom.14} we have analyzed the evolution of 
quadrupole and octupole shapes using a consistent microscopic framework based on 
relativistic EDFs. In thorium isotopes, in particular, the calculated triaxial quadrupole and 
axial quadrupole-octupole energy surfaces, and predicted 
observables (excitation energies, isotope shifts of charge radii, electromagnetic
transition rates) point to the occurrence of a simultaneous phase
transition between spherical and quadrupole-deformed prolate shapes,
and between non-octupole and octupole-deformed  shapes, with $^{224}$Th being
closest to the critical point of the double shape phase transition \cite{Li.13}. 

%----------------------------------------------------------------------------------------------------------------
\begin{figure}[]
\begin{center}
\begin{tabular}{ccc}
\includegraphics[scale=0.275]{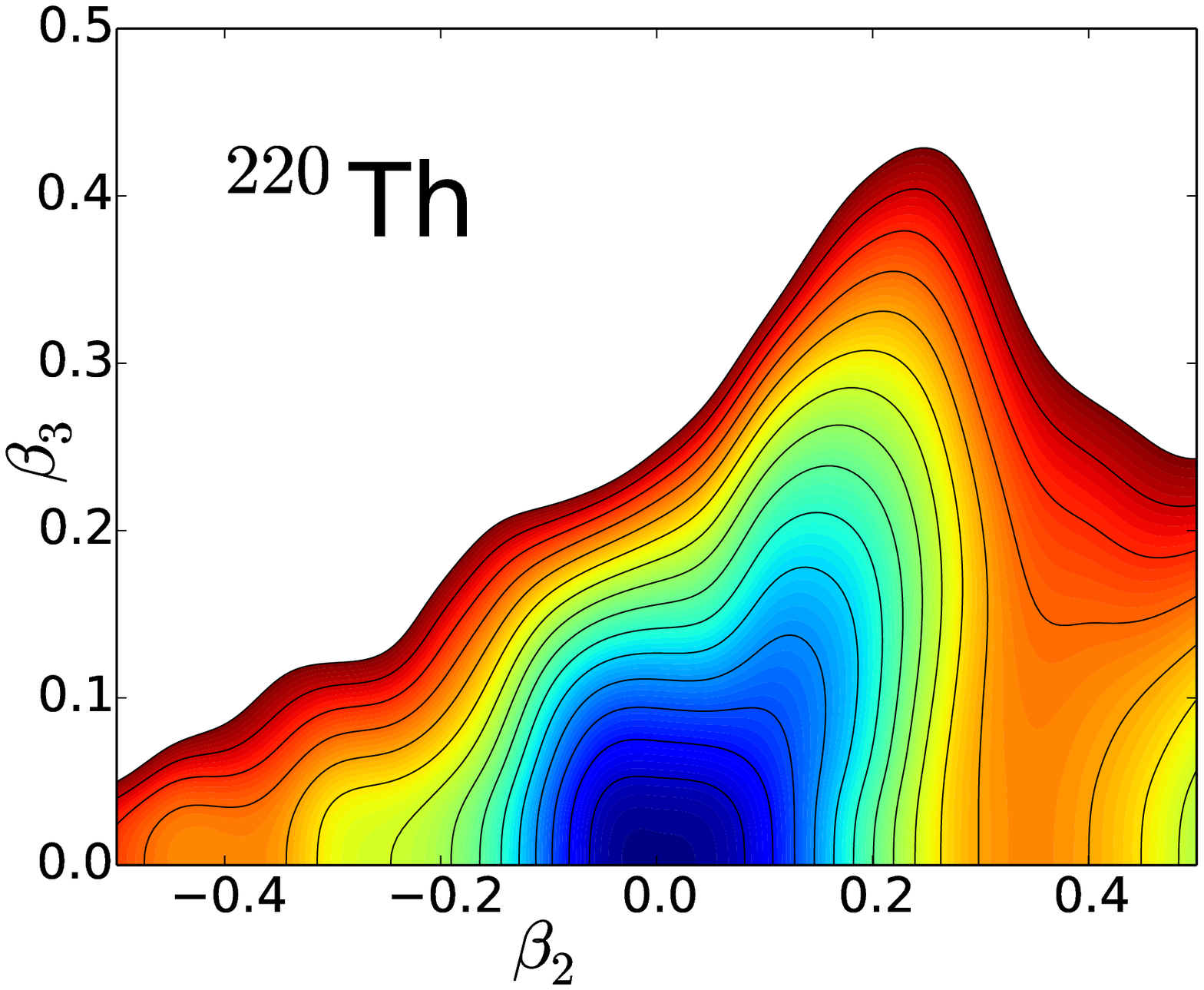}
\includegraphics[scale=0.275]{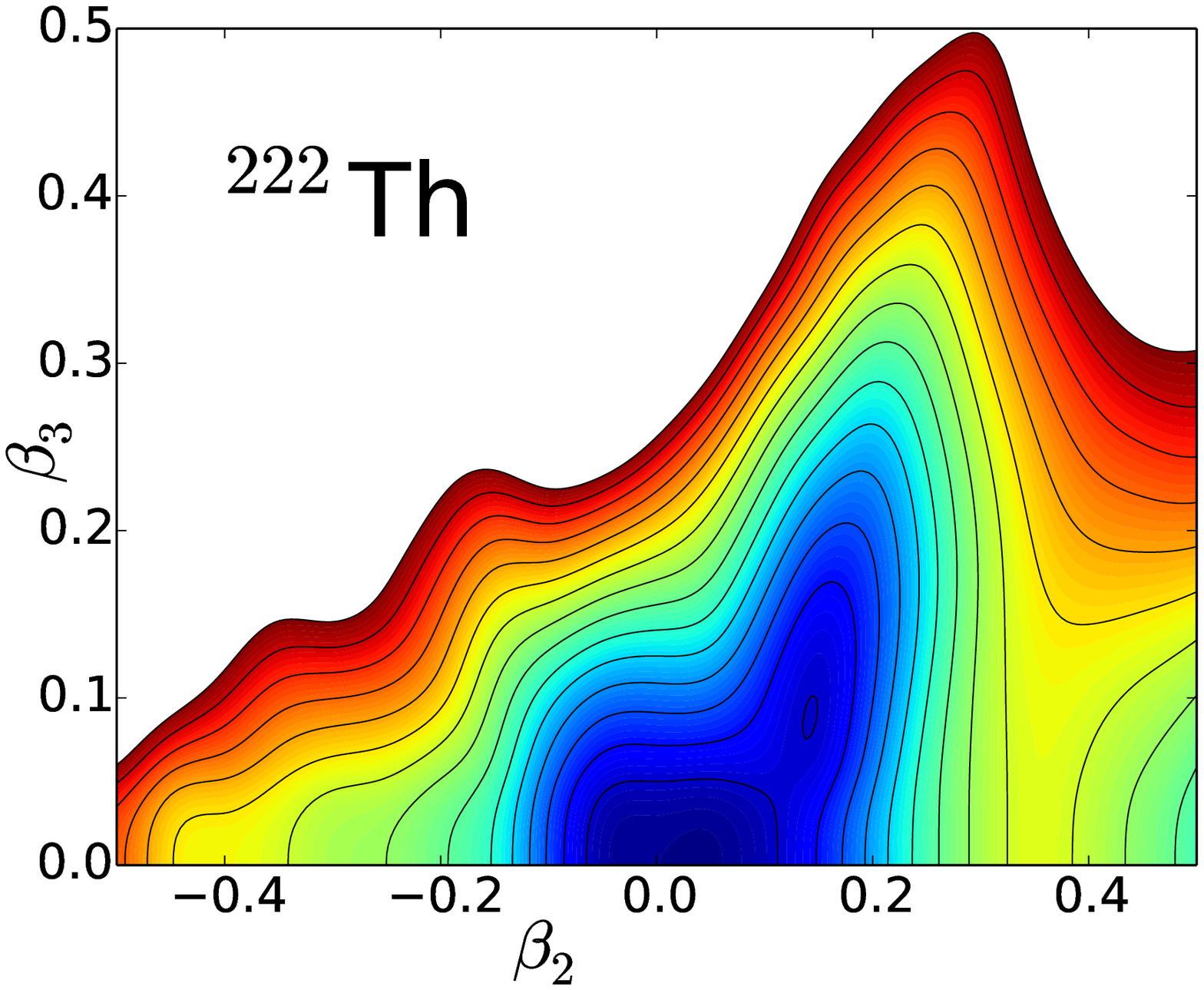}
\includegraphics[scale=0.275]{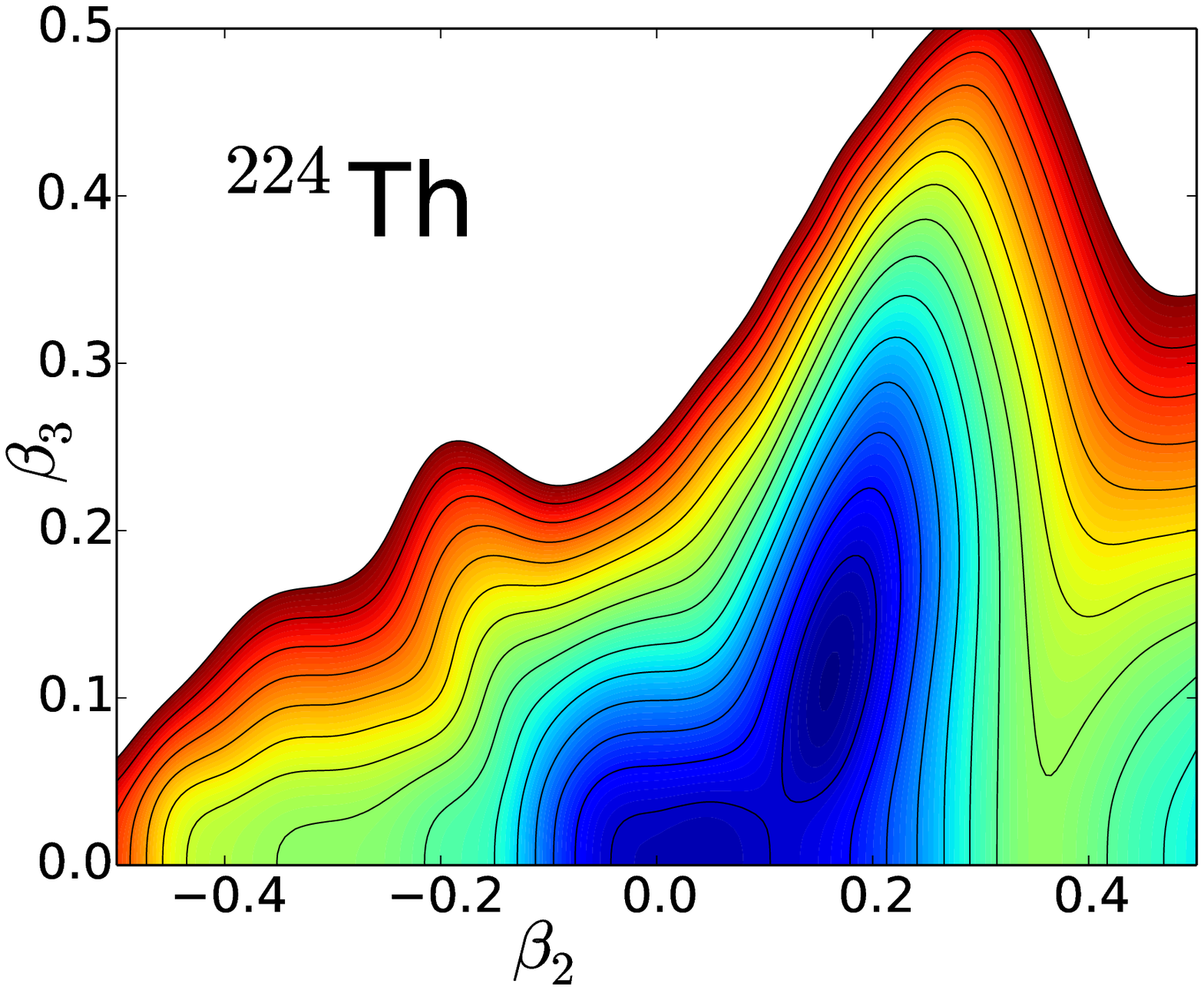} \\
\includegraphics[scale=0.275]{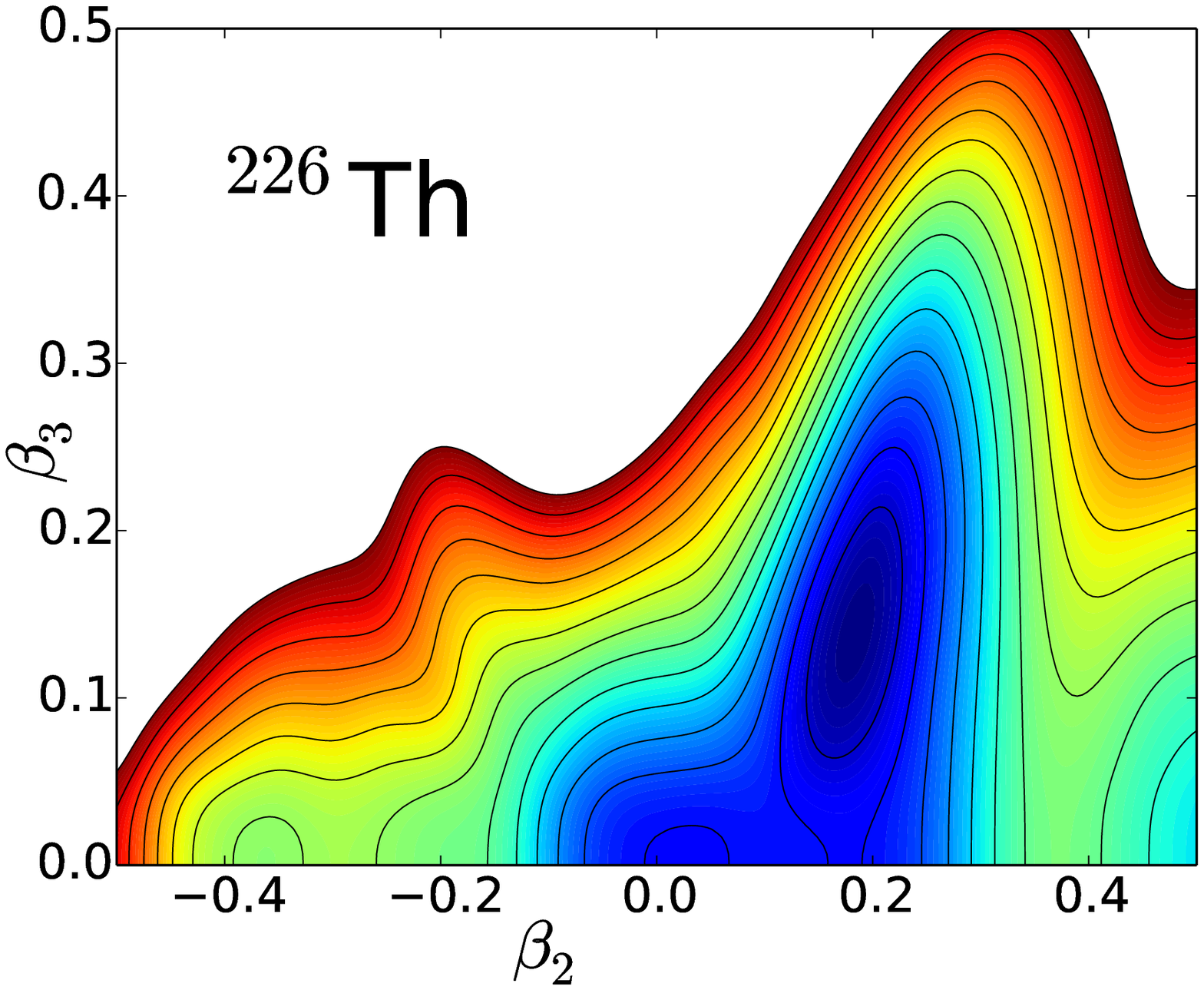}
\includegraphics[scale=0.275]{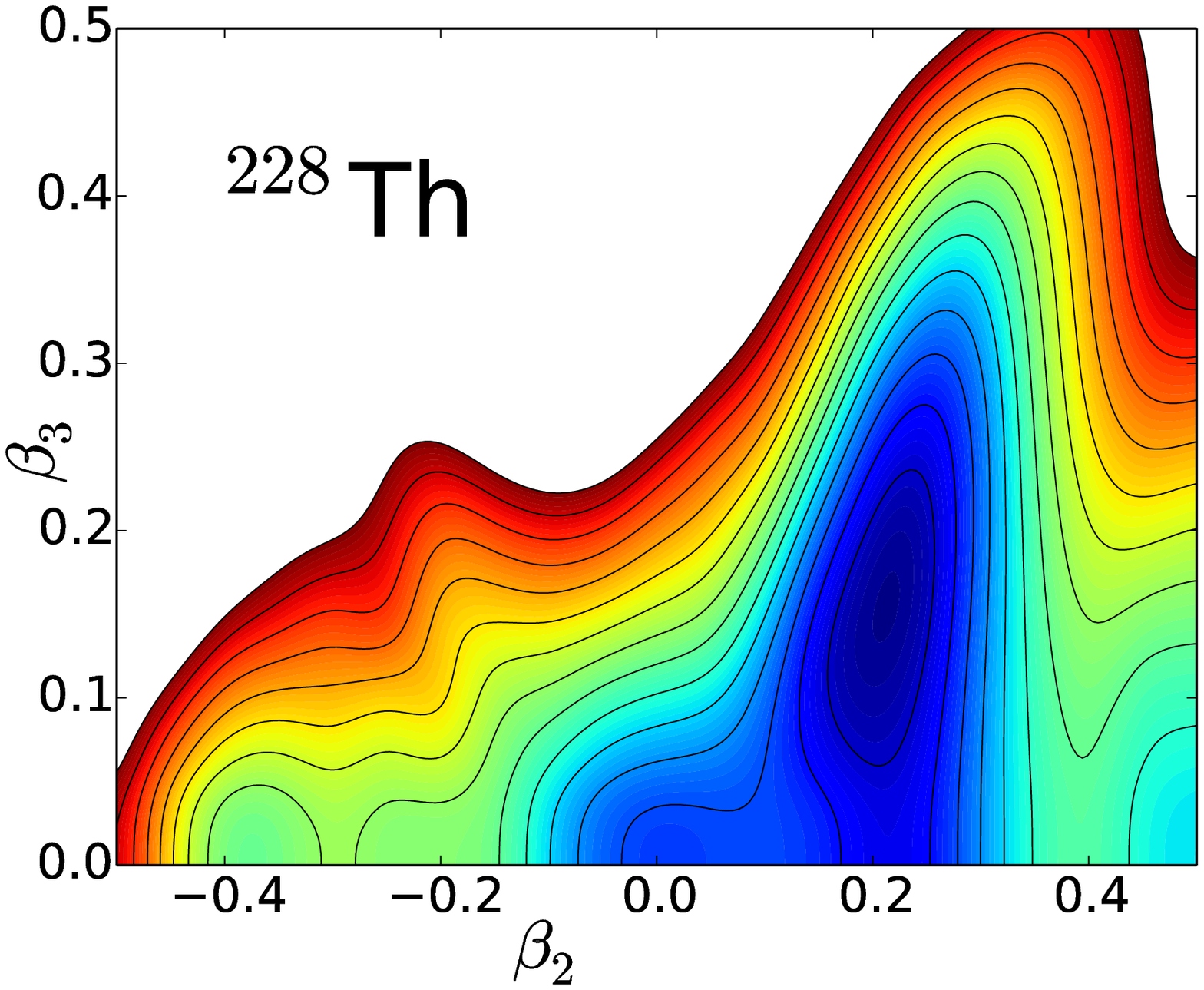}
\includegraphics[scale=0.275]{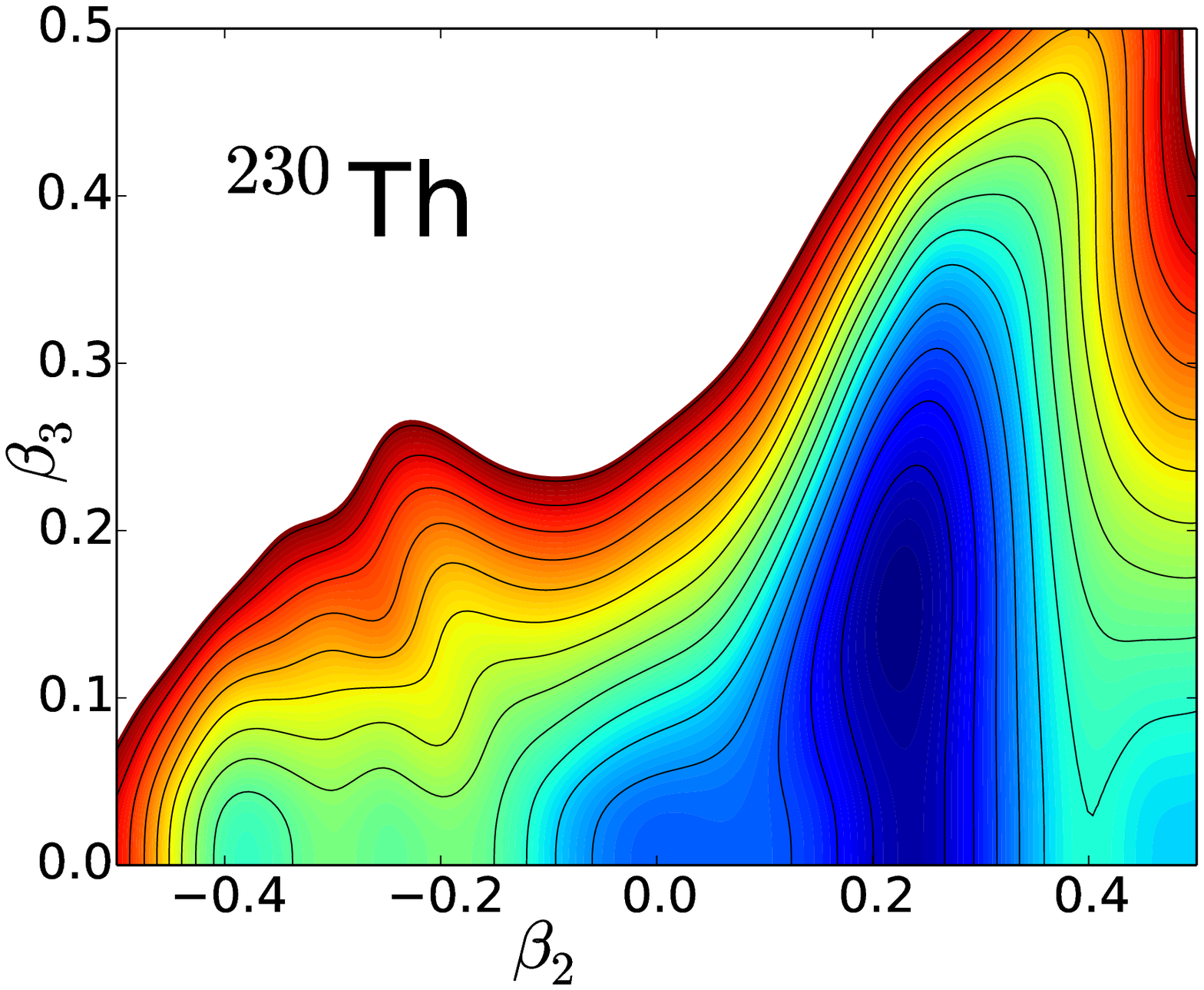}
\end{tabular}
\caption{\label{fig:th_pes} Axially symmetric energy surfaces of the isotopes
 $^{220-230}$Th in the $(\beta_{2},\beta_{3})$ 
 plane. The contours join points on the surface with the same energy
 and the energy difference between neighboring contours is
 1 MeV. Positive (negative) values of $\beta_{2}$ correspond to prolate (oblate) 
 configurations. Energy surfaces are symmetric with respect to the
 $\beta_{3}=0$ axis.}
\end{center}
\end{figure}
%----------------------------------------------------------------------------------------------------------------

Figure~\ref{fig:th_pes} displays the deformation energy surfaces in the plane 
of axial quadrupole and octupole deformation parameters 
($\beta_{2}$, $\beta_{3}$) for the isotopes $^{220 - 230}$Th. 
This isotopic chain exhibits an interesting structural evolution, visible already 
at the SCMF level. A rather soft energy surface is calculated for $^{220}$Th with the 
minimum at $(\beta_{2}, \beta_{3})\approx (0, 0)$, and this will give rise to quadrupole 
vibrational excitation spectra. Quadrupole deformation becomes more pronounced in $^{224}$Th, 
and one also notices 
the emergence of octupole deformation. The energy minimum is found in the 
$\beta_{3}\neq 0$ region, located at $(\beta_{2}, \beta_{3})\approx (0.1, 0.1)$ . 
From $^{224}$Th to $^{228}$Th the occurrence of a rather strongly marked 
octupole minimum is predicted. Starting from $^{228}$Th, the minimum 
becomes softer in the octupole $\beta_{3}$ direction. An octupole-soft surface, 
almost completely flat in $\beta_{3}$ for $\beta_{3} \leq 0.3$,
is calculated for $^{232}$Th. 

%----------------------------------------------------------------------------------------------------------------
\begin{figure}
\begin{center}
\includegraphics[scale=0.6]{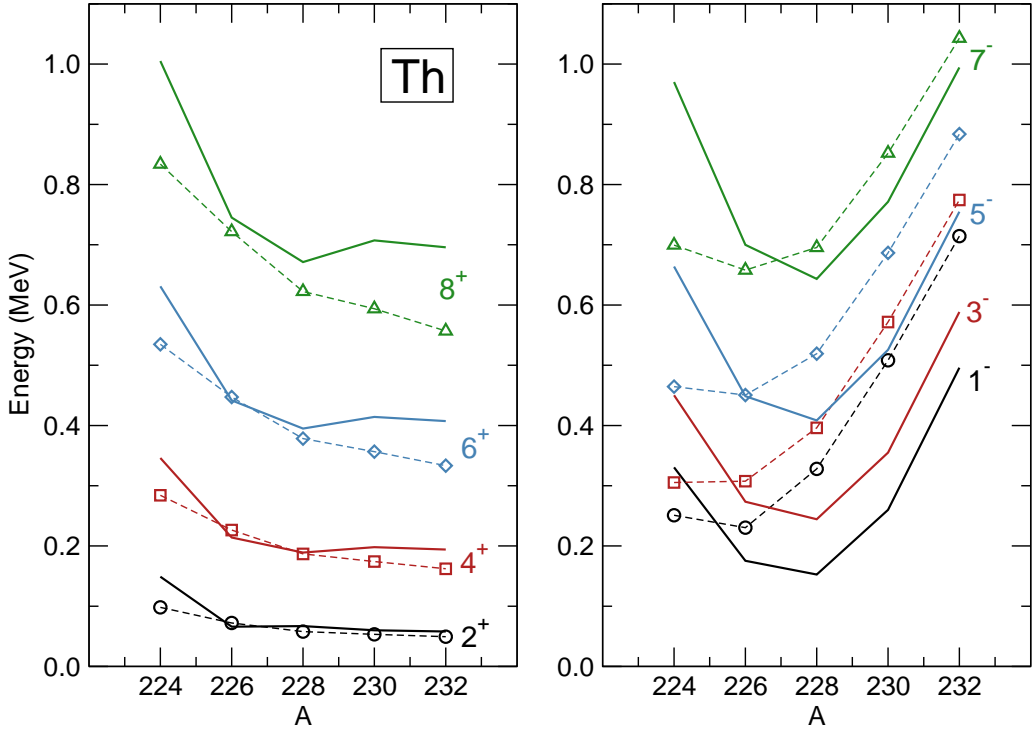}
\caption{\label{fig:th_spec}Excitation energies of low-lying yrast positive-parity (left)
and negative-parity (right) collective states of $^{224-232}$Th. Lines and symbols denote the 
 theoretical and experimental 
 \cite{data} levels, respectively.}
\end{center}
\end{figure}
%----------------------------------------------------------------------------------------------------------------

In Fig.~\ref{fig:th_spec} we analyse the systematics of energy spectra of the positive-parity 
ground-state band ($K^{\pi}=0^{+}$) (left) and the lowest negative-parity ($K^\pi =0^-$) sequences 
(right) in $^{224-232}$Th. The theoretical values calculated using the quadrupole-octupole collective 
Hamiltonian~(\ref{eq:CH2}) are shown in comparison to available data \cite{data}.  The excitation energies of 
positive-parity states systematically decrease with mass number, reflecting the increase of 
quadrupole collectivity. $^{220,222}$Th exhibit a quadrupole vibrational structure, whereas 
pronounced ground-state rotational bands with
$R_{4/2}=E(4^{+}_{1})/E(2^{+}_{1})\approx 3.33$ are calculated in
$^{226-232}$Th. For the lowest negative-parity bands
the excitation energies display a parabolic 
structure centered between $^{224}$Th and $^{226}$Th. 
The approximate parabola of $1^{-}_{1}$ states has a minimum 
at $^{226}$Th, in which the octupole deformed minimum 
is most pronounced. Starting from $^{226}$Th the energies of negative-parity 
states systematically increase and the band becomes more 
compressed. A rotational-like collective band based on the octupole vibrational
state, i.e., the $1^{-}_{1}$ band-head, develops. The parabolas of negative-parity 
states calculated with the quadrupole-octupole Hamiltonian are in qualitative agreement 
with data, although the minima are predicted to occur at $^{228}$Th rather than $^{226}$Th. 
Note, however, that all levels shown in Fig.~\ref{fig:th_spec} are below 1 MeV excitation 
energy, so that the differences between calculated and experimental levels are rather 
small, especially considering that no parameters were adjusted to data. The approximations 
involved in the calculation of the Hamiltonian parameters (perturbative cranking for the 
mass parameters and the Inglis-Belyaev formula for the moments of inertia) determine 
the level of quantitative agreement with experiment. We note that the theoretical 
B(E2) values for transitions within the ground state bands are in agreement 
with availabe data, while the calculated $B(E3; 3^-_1\to 0^+_1)$: 
61 W.u. for $^{230}$Th and 41 W.u. for $^{232}$Th, are somewhat larger than 
the experimental values: 29(3) W.u. for $^{230}$Th and 24(3) W.u. for $^{232}$Th \cite{Kib.02}. 

%----------------------------------------------------------------------------------------------------------------
\begin{figure}
\begin{center}
\includegraphics[scale=0.6]{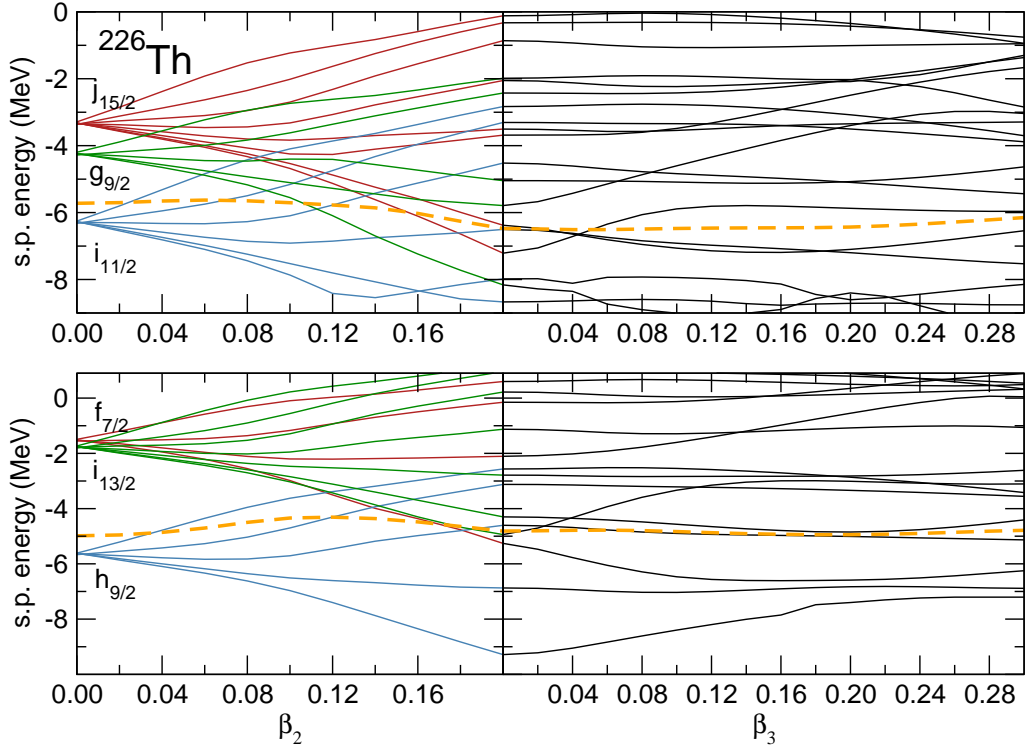}
\caption{\label{fig:th226-single-particle} $^{226}$Th single-neutron (upper panel) and single-proton
(lower panel) levels in the canonical basis as functions of the deformation
parameters. The path follows the quadrupole deformation parameter
$\beta_2$ up to the position of the equilibrium minimum $\beta_2=0.2$, with 
the constant octupole deformation $\beta_3=0$ (left panels). For $\beta_2=0.2$ 
the panels on the right display the single-nucleon energies from $\beta_3=0$ to $\beta_3=0.3$. 
Dashed curves denote the position of the Fermi level at each deformation.}
\end{center}
\end{figure}
%----------------------------------------------------------------------------------------------------------------

Let us consider in more detail the structure of $^{226}$Th which, at the SCMF level, exhibits a nice 
example of coexistence of axial quadrupole and octupole minima. The energy 
difference between the local minimum at $\beta_3 = 0$ and the equlibrium minimum 
at $\beta_3 = 0.16$ is 1.8 MeV. We have also carried out a constrained triaxial 
quadrupole SCMF calculation which confirms that the quadrupole minimum is indeed at 
$\gamma = 0^\circ$, that is, axial prolate. The microscopic origin of coexistence 
of the two minima becomes apparent from the dependence of the single-nucleon
levels on the two deformation parameters. Figure \ref{fig:th226-single-particle}
displays the single neutron and proton levels
of $^{226}$Th along a path in the $\beta_2 - \beta_3$ plane. Starting from the spherical configuration,
the path follows the quadrupole deformation parameter $\beta_2$
up to the position of the equilibrium minimum $\beta_2 = 0.2$, with the
octupole deformation parameter kept constant at zero value. Then,
for the constant value $\beta_2 = 0.2$, the path continues from
$\beta_3 = 0$ to $\beta_3 = 0.3$. The necessary condition for the occurrence of 
low-energy octupole collectivity is the presence of pairs of orbitals near the Fermi level 
that are strongly coupled by the octupole interaction. In the panels on the left of Fig.~\ref{fig:th226-single-particle} 
we notice states of opposite parity that originate from the spherical levels $g_{9/2}$ and $j_{15/2}$ for neutrons, 
and $f_{7/2}$ and $i_{13/2}$ for protons. The total energy can be related to the level density 
around the Fermi surface, that is, a lower-than-average density of single-particle levels results in extra 
binding. Therefore, the local quadrupole minimum seen on the axial energy surface 
of $^{226}$Th reflects the $\beta_2$-dependence of the levels of the Nilsson diagram 
for $\beta_3 = 0$.  
For the levels in the panels on the right of Fig.~\ref{fig:th226-single-particle} 
parity is not conserved, and the only quantum number that 
characterises these states is the projection of the angular momentum on the symmetry axis. 
The octupole minimum, rather soft along the $\beta_3$-path in $^{226}$Th, 
is attributed to the low density of both proton and neutron states close to the 
corresponding Fermi levels in the interval of deformations around $\beta_3 = 0.16$.

Fully microscopic analyses of coexistence of quadrupole and octupole shapes and, in general, the 
evolution of octupole correlations in heavy nuclei, will be particularly important for future experimental  
studies of reflection-asymmetric shapes using accelerated radioactive beams \cite{Gaffney.13},  
and in searches for new symmetry violating interactions beyond the standard model \cite{BW.15}. 
%============================================================
%  Section 6
\section{\label{secVI}Summary}
%============================================================

The framework of nuclear energy density functionals provides an intuitive and yet accurate 
microscopic interpretation of the evolution of single-nucleon shell structure and the related 
phenomena of deformations, shape transitions and shape coexistence. Self-consistent mean-field 
calculations of deformation energy surfaces produce symmetry-breaking many-body states 
that include important static correlations such as deformations and pairing. Intrinsic 
shapes that correspond to minima on the deformation energy surface are determined by the 
static (constrained) collective coordinates. Dynamical correlations are included by  
collective structure models that restore symmetries broken by the static mean field 
and take into account quantum fluctuations of collective variables.
The microscopic input 
for the generator coordinate method or the collective Hamiltonian model are completely 
determined by the choice of the energy density functional and pairing interaction. These models 
can be used to calculate observables that characterise the evolution and eventual 
coexistence of different shapes: low-energy excitation spectra, electromagnetic transition rates, 
changes in masses (separation energies), isotope and isomer shifts, and that can be directly 
compared to data.  

Using a single relativistic energy density functional and a finite-range pairing interaction 
separable in momentum space, in the present study we have analysed illustrative examples 
of diverse phenomena related to evolution of shell structure: shape transition and 
coexistence in neutron-rich $N=28$ isotones, the structure of
lowest $0^+$ excitations in deformed $N = 90$ rare-earth nuclei, %$N \approx 90$ rare-earth nuclei, 
and quadrupole and octupole shape transitions 
in thorium isotopes. Spectroscopic properties have been calculated using the 
five-dimensional quadrupole and axial quadrupole-octupole collective Hamiltonian models. 
The very good agreement between theoretical predictions and available data, and especially 
the fact that very different regions of the mass table could be considered with no 
need to adjust or fine-tune model parameters to specific data, demonstrate that the approach 
based on universal density functionals is a method of choice for studies of shape transitions 
and coexistence over the entire table of nuclides, including regions of exotic short-lived 
nuclei far from stability.  

Future developments of structure methods based on nuclear energy density functionals 
include a number of major challenges. One of the most important and certainly most 
difficult is the construction of a consistent set of approximations for the exchange-correlation 
energy functional. In the context of phenomena discussed in the present analysis, 
it would be very interesting to try to develop microscopic functionals that, in addition to the 
dependence on ground-state densities and currents composed of occupied Kohn-Sham 
orbitals, also include a dependence on unoccupied orbitals. This is particularly important 
for studies of the evolution of shell structure and modification of gaps in nuclei far from 
stability and/or superheavy nuclei. For quantitative comparison with available 
spectroscopic data and predictions in new regions of the chart of nuclides, 
accurate and efficient algorithms have to be developed that perform a complete 
restoration of symmetries broken by self-consistent mean-field solutions for general 
quadrupole and octupole shapes. Deformation-dependent parameters of  
collective quadrupole (-octupole) Hamiltonians (vibrational inertial functions and moments 
of inertia) need to be determined using methods that go beyond the simple cranking 
formulas and include the full dynamics of a nuclear system. Finally, theoretical studies 
of low-energy spectroscopic properties that characterise shape coexistence will also have to 
include uncertainty estimates, quantify theoretical errors and evaluate correlations 
between observables.

%\begin{acknowledgements}
\ack{
The authors acknowledge discussions with Bing-Nan Lu,  P. Marevi\'{c}, K. Nomura, P. Ring and J. M. Yao.
This work was supported in part by the Chinese-Croatian project "Universal models of exotic
nuclear structure", the Croatian Science Foundation under the project "Structure and Dynamics
of Exotic Femtosystems" (IP-2014-09-9159), and the NSFC project No. 11475140.}
%\end{acknowledgements}
\bigskip
\bigskip
%\section*{References}
%\bibliographystyle{unsrt}
%\bibliography{Coexistence}

\begin{thebibliography}{00}

\bibitem{HW.11} K. Heyde and J. L. Wood, Rev. Mod. Phys. 83, 1467 (2011). 

\bibitem{BN.96} P. A. Butler and W. Nazarewicz, Rev. Mod. Phys. 68, 349 (1996)

\bibitem{SP.08} O. Sorlin and M.-G. Porquet, Prog. Part. Nucl. Phys. 61, 602 (2008).

\bibitem{CJC.10} P. Cejnar, J. Jolie, and R. F. Casten, Rev. Mod. Phys. 82, 2155 (2010).

\bibitem{BHR.03} M. Bender, P.-H. Heenen, and P.-G. Reinhard, 
	Rev. Mod. Phys. 75, 121 (2003).
	
\bibitem{VALR.05} 
 	D. Vretenar, A.V. Afanasjev, G.A. Lalazissis, and P. Ring, 
	Phys. Rep. \textbf{409}, 101 (2005).
	
\bibitem{LNP.641} 
G. A. Lalazissis, P. Ring, and D. Vretenar (Eds.), 
	\textit{Extended Density Functionals in Nuclear Structure Physics},
	Lecture Notes in Physics \textbf{641}, (Springer, Heidelberg 2004.)  

\bibitem{SR.07} J. R. Stone and P.-G. Reinhard, Prog. Part. Nucl. Phys. 58, 587 (2007).

\bibitem{Laughlin.00} R. B. Laughlin, D. Pines, J. Schmalian, B. P. Stojkovi\' c, and P. Wolynes, 
	Proc. Natl. Acad. Sci. 97, 32 (2000).

\bibitem{NVR.11} T. Nik\v si\' c, 	D. Vretenar, and P. Ring, Prog. Part. Nucl. Phys. 66, 519 (2011).
	
\bibitem{LNP.879} T. Duguet, Lecture Notes in Physics \textbf{879}, (Springer, Heidelberg 2014) p. 293

\bibitem{RS.80} P. Ring and P. Schuck, {\em The Nuclear Many-Body Problem}
	(Springer-Verlag, Heidelberg, 1980).
	
\bibitem{BH.08} M. Bender and P.-H. Heenen, Phys. Rev. C 78, 024309 (2008).

\bibitem{RE.10}  T. R. Rodr\' iguez and J. L. Egido, Phys. Rev. C 81, 064323 (2010).

\bibitem{Yao.10} J. M. Yao, J. Meng, P. Ring, and D. Vretenar,
    Phys. Rev. C 81, 044311 (2010).
	
\bibitem{BABH.14} B. Bally, B. Avez, M. Bender, and P.-H. Heenen, Phys. Rev. Lett. 113, 162501 (2014). 

\bibitem{RG.87}
P.-G. Reinhard and K. Goeke, Rep. Prog. Phys. {50},  1  (1987).

\bibitem{PR.09} L. Pr{\'o}chniak and S. G. Rohozi{\' n}ski,
	J. Phys. {G 36},  123101 (2009).

\bibitem{GR.80} K. Goeke and P.-G. Reinhard, Ann. Phys. (N.Y.) {124},  249 (1980).

 \bibitem{Delaroche10} J. -P. Delaroche, M. Girod, J. Libert, H. Goutte, S. Hilaire, S. P\'{e}ru, 
	N. Pillet, and G. F. Bertsch, Phys. Rev. C {\bf 81}, 014303 (2010).
	
\bibitem{Nik.08} T. Nik\v{s}i\'{c}, D. Vretenar, P. Ring, Phys. Rev. C 78, 034318  (2008).

\bibitem{Tian_PLB.09} Y. Tian, Z.~Y. Ma, P. Ring, Phys. Lett. B 676, 44 (2009).

\bibitem{Yao.14} J. M. Yao, K. Hagino, Z. P. Li, J. Meng, and P. Ring
		Phys. Rev. C 89, 054306 (2014).

\bibitem{Nik.09}  T. Nik\v{s}i\'{c}, Z. P. Li, D. Vretenar,  L. Pr{\'o}chniak,
	J. Meng, and P. Ring, Phys. Rev. C 79, 034303 (2009).
		
\bibitem{Hin.12} N. Hinohara, Z. P. Li, T. Nakatsukasa, T. Nik\v{s}i\'{c}, and D. Vretenar
	Phys. Rev. C 85, 024323 (2012).

\bibitem{GG.79} M. Girod and B. Grammaticos, Nucl. Phys. {A 330},  40  (1979).

 \bibitem{Li.11} Z. P. Li, J. M. Yao, D. Vretenar, T. Nik\v{s}i\'{c}, H. Chen, and J. Meng,
   Phys. Rev. C 84, 054304 (2011).
   
\bibitem{RE.11} T. R. Rodr\'iguez, J. L. Egido, Phys. Rev. C 84, 051307 (R) (2011).
  
 \bibitem{Grevy.05} S. Gr\'evy {\it et al.}, Eur. Phys. J. A 25, 111 (2005). 
 
 \bibitem{Glasmacher.97} T. Glasmacher {\it et al.}, Phys. Lett. B 395, 163 (1997).
 
 \bibitem{Force.11} C. Force {\it et al.}, Phys. Rev. Lett. 105, 102501 (2011).

 \bibitem{Chevrier.14} R. Chevrier, L. Gaudefroy, Phys. Rev. C 89, 051301(R) (2014).     

 \bibitem{Santiago-Gonzalez.11} D. Santiago-Gonzalez {\it et al.}, Phys. Rev. C 83, 061305 (2011).
 
  \bibitem{Utsuno.15} Y. Utsuno, N. Shimizu, T. Otsuka, T. Yoshida, Y. Tsunoda,
   Phys. Rev. Lett. 114, 032501 (2015).
   
  \bibitem{Sorlin.10} O. Sorlin, Nucl. Phys. A 834, 400c (2010).
  
\bibitem{CMC.07} R. F. Casten and E. A. McCutchan, J. Phys. G: Nucl. Part. Phys. 34, R285 (2007).    

\bibitem{NVL.07} T. Nik\v{s}i\'{c}, D. Vretenar, G. A. Lalazissis, and P. Ring, 
	Phys. Rev. Lett. 99, 092502 (2007). 
	
\bibitem{LNV.09} Z. P. Li, T. Nik\v{s}i\'{c}, D. Vretenar, J. Meng, G. A. Lalazissis, and P. Ring, 
	Phys. Rev. C 79, 054301 (2009).
	
\bibitem{Li.09} Z. P. Li, T. Nik\v{s}i\'{c}, D. Vretenar, and J. Meng,
	Phys. Rev. C 80, 061301(R) (2009).
	
\bibitem{data} Brookhaven National Nuclear Data Center, http://www.nndc.bnl.gov.

\bibitem{Garr.01} P. E. Garrett, J. Phys. G. 27, R1 (2001). 

\bibitem{DGL.10} J. -P. Delaroche, M. Girod, J. Libert, H. Goutte, S. Hilaire, S. P\' eru, N. Pillet, and G. F. Bertsch,
		Phys. Rev. C 81, 014303 (2010).
		
\bibitem{Kulp.08} W. D. Kulp {\it et al.}, Phys. Rev. C 77, 061301 (R) (2008). 

\bibitem{Garr.09} P. E. Garrett {\it et al.}, Phys. Rev. Lett. 103, 062501 (2009).

\bibitem{PP.93} S. Pilat and K. Pomorski, Nucl. Phys. A 554, 413 (1993).

\bibitem{Ber.11} R. Bernard, H. Goutte, D. Gogny, and W. Younes, Phys. Rev. C 84, 044308 (2011).	

\bibitem{Kulp.03} W. D. Kulp {\it et al.}, Phys. Rev. Lett. 91, 102501 (2003).

\bibitem{Kulp.05} W. D. Kulp {\it et al.}, Phys. Rev. C 71, 041303 (R) (2005).
		
\bibitem{BW.15} P. A. Butler and L. Willmann, Nucl. Phys. News 25, 12 (2015).

\bibitem{Nom.13} K. Nomura, D. Vretenar, and Bing-Nan Lu, Phys. Rev. C 88, 021303(R) (2013).

\bibitem{Li.13} Z. P. Li, B. Y. Song, J. M. Yao, D. Vretenar, and J. Meng,  Phys. Lett. B 726, 866 (2013).

\bibitem{Nom.14} K. Nomura, D. Vretenar, T. Nik\v{s}i\'{c}, and Bing-Nan Lu, Phys. Rev. C 89, 024312 (2014).

\bibitem{Kib.02} T. Kibedi, R.H. Spear, At. Data Nucl. Data Tables 80, 35(2002).

\bibitem{Gaffney.13} L. P. Gaffney et al., Nature 497, 199 (2013). 


\end{thebibliography}

\end{document}